\documentclass{aa}
\usepackage{natbib}
\bibpunct{(}{)}{;}{a}{}{,}
\usepackage{amsfonts}
\usepackage{psfrag}
\usepackage{graphicx}

\newcommand{\Msolar}{\mbox{\,$\rm M_{\odot}$}}        
\begin{document}
\title{3D kinematics of white dwarfs from the SPY project
\thanks{Based on observations obtained at the Paranal Observatory of the European Southern Observatory for programs 165.H-0588 and 167.D-0407}}
\author{E.-M. Pauli \inst{1} 
\and R.Napiwotzki \inst{1} 
\and M. Altmann \inst{1}   
\and U. Heber \inst{1}
\and M. Odenkirchen \inst{2} 
\and F. Kerber \inst{3}
} 
\institute{Dr.~Remeis-Sternwarte, Astronom.\ Institut, Universit\"at 
        Erlangen-N\"urnberg, Sternwartstr.~7, 96049 Bamberg, Germany
\and
Max-Planck-Institut f\"ur Astronomie, K\"onigstuhl 17, 69117 Heidelberg, 
 Germany
\and 
Space Telescope European Coordinating Facility, ESO, Karl-Schwarzschild-Str.~2, 85748 Garching, 
        Germany
}
\date{}
\offprints{E.-M. Pauli (pauli@sternwarte.uni-erlangen.de)}

\abstract{

We present kinematics of a sample of 107 DA
white dwarfs from the SPY project (ESO SN Ia Progenitor surveY)  
and discuss kinematic criteria for a distinction of thin disk, 
thick disk, and halo populations.
This is the first homogeneous sample of white dwarfs for which 3D
space motions have been determined. 

Since the percentage of old stars among white dwarfs is higher than 
among main-sequence stars, they are presumably valuable tools 
in studies of old populations such as the halo and the thick disk. 
Studies of white dwarf kinematics can help to determine the fraction 
of the total mass of our Galaxy contained in 
the form of thick disk and halo white dwarfs, an issue which 
is still under discussion.  
 
Radial velocities and spectroscopic distances obtained by the 
SPY project are combined with our measurements of proper motions 
to derive 3D space motions. 
Galactic orbits and further kinematic parameters are computed. 
We calculate individual errors of kinematic parameters by means of a  
Monte Carlo error propagation code.
Our kinematic criteria for assigning population membership are     
deduced from a sample of F and G stars taken from the literature 
for which chemical criteria can be used to distinguish 
between thin disk, thick disk and halo. 

Candidates for thick disk and halo members are selected 
in a first step from the classical $U$\/-$V$-velocity diagram. 
Our final assignment of population membership is based on orbits 
and position in the $J_{\mathrm{z}}$-eccentricity diagram.
We find four halo and twelve thick disk white dwarfs.  
 
We also present a systematic study of the effects of ignoring  
the radial velocity in kinematic investigations. 

}

\maketitle
\section{Introduction: population membership of white dwarfs\label{intro}} 

White dwarfs are the evolutionary end-products of most stars. 
As they are faint objects only the nearby objects have been detected so far. 
However, a large number of white dwarfs should be present in the Galaxy. 
Determining the contribution of white dwarfs to the 
total mass of the Galaxy 
could help solve one of the fundamental questions in 
modern astronomy: what is the nature of dark matter? 
The fact that the rotation curves of many galaxies are not Keplerian \citep{rubin78} 
invokes the existence of additional dark matter distributed in 
a near-spherical structure, the so-called heavy-halo \citep{ostriker73}.
It is estimated that for the Milky Way only 10\% of the total 
mass are present in the form of stars, gas, and dust in the Galactic disk 
and halo \citep{alcock00}.   
Dark matter candidates for the remaining 90\% include exotic particles, 
cold molecular gas, and compact objects like black holes, 
white dwarfs and brown dwarfs. 
The role of white dwarfs in the dark matter problem is still uncertain.  
An open issue is the fraction of white dwarfs in the 
thick disk and halo populations and their 
fraction of the total mass of the Galaxy.
In this context kinematic studies have proved a useful 
tool to decide on population membership of white dwarfs.  

\citet{sion88} were the first to carry out kinematic investigations of a 
large sample of white dwarfs from the first edition of the McCook \& Sion catalogue 
\citep{mccook87}. They found a fraction of about $10$\% halo 
white dwarfs.

Another method for a distinction of the disk and halo populations 
was proposed by 
\citet{garcia99}: they developed a neural network 
to classify a subsample of the McCook \& Sion stars \citep{mccook87} 
in a five dimensional parameter 
space by means of synthetic halo and disk tracer stars generated with a Monte 
Carlo simulation. Their results confirm those of  
\citet{sion88}.

\citet{liebert89} investigated a sample of 43 
spectroscopically confirmed white dwarfs from the LHS catalogue 
\citep{luyten79}.  
They obtained a percentage of 14\% halo white dwarfs.   
They were also the first to derive a luminosity function of halo 
white dwarfs.

Recently \citet{oppenheimer01} have completed a 
deep proper motion survey towards the 
South Galactic Cap, 
and claim detection of 38 halo white dwarfs.
Most of the white dwarfs have featureless DC spectra 
and therefore radial velocities could not be measured. 
The Galactic radial ($U$) and rotational ($V$) velocity components 
were calculated from photometric distances and proper motions alone. 
The velocity component $W$ perpendicular to the Galactic disk was set 
to zero arguing that in the  direction of the 
South Galactic Cap the tangential motion is not a function of $W$. 
Their results are presented in the form of a $U$\/-$V$-velocity diagram 
with superposed $1\sigma$ and $2\sigma$ contours for the expected locations 
of the thick disk and halo components of the Galaxy. 
White dwarfs lying outside the $2\sigma$ contours of the disk are 
assumed to belong to the heavy halo. 

However, this result is discussed controversially: e.g.\ \citet{reid01} 
claimed that the velocity distribution of the so-called heavy 
halo white dwarfs 
is more consistent with the high velocity tail of the thick disk. 
The major difference between the two investigations is that 
\citet{oppenheimer01} 
adopted $W=0\,\rm{km\,s^{-1}}$ whereas \citet{reid01} set $v_{\rm rad}=0\,\rm{km\,s^{-1}}$.    

The common problem of the investigations discussed above is the lack 
of radial velocity measurements. Especially deviating conclusions derived 
from the white dwarfs of the \citet{oppenheimer01} sample 
demonstrate that different assumptions on the values of $v_{\rm rad}$ 
can produce different fractions of halo and thick disk stars 
and thus have effects on the determination 
of the white dwarf halo density. 
Therefore a sample of white dwarfs with known radial velocity measurements
 is needed in 
order to obtain the full 3D kinematic information and to study the effects
of setting $v_{\rm rad}$ to zero. 
Once this effect is well understood samples of stars can be dealt with 
for which no radial velocity information is available, 
e.g.\ in the case of cool white dwarfs. 

\citet{silvestri01} presented kinematics of 41 
white dwarfs in common proper 
motion binary systems. Radial velocities 
could be easily measured from the sharp lines in the spectrum of the cool 
companion star. 
Thus they could calculate 
all three velocity components $U$, $V$, and $W$. The mean values and standard 
deviations of those indicate that most of the sample stars belong to the 
thick disk. Three white dwarfs (7\%) were found to be belonging to the halo. 

In a more recent study, \citet{silvestri02} investigated 
kinematics of another 116 white
dwarfs with M dwarf companions. They detected 13 high velocity white
dwarfs and concluded that 12 of them belong to the thick disk as their
M dwarf companions have near solar abundance levels. 
They compared the effect of assuming either $v_{\rm rad}$ or $W$ to be zero and
discovered that the effect of $v_{\rm rad}=0\,\rm{km\,s^{-1}}$ is negligible, whereas $W=0$
changes the fraction of high velocity white dwarfs. 

We present a sample of DA white dwarfs from the ESO {\bf S}N Ia 
{\bf P}rogenitor surve{\bf Y} (SPY) by \citet{napiwotzki01} 
which is ideal for probing population membership of white dwarfs. 
The SPY sample allows us to overcome several limitations of 
previous investigations.
When investigating DA white dwarfs, radial velocities can be measured from 
the shifts of the Balmer lines which is an advantage over DC 
white dwarfs where no spectral lines are present.
Due to high resolution UVES VLT spectra 
we can benefit from radial velocities of unprecedented precision 
(errors of only $2\,{\rm km\,s^{-1}}$) and spectroscopic distances  
(relative errors of only 10\%) from Napiwotzki et al. (2003, in prep.).  
We supplemented these data with proper motion measurements 
of fair quality (typical errors about $20\,\rm{mas\,yr^{-1}}$).
Therefore we possess a very homogeneous set of radial and tangential velocity 
information with individual errors for each star. 
Contrary to previous studies we do not only consider the classical velocity 
components $U$, $V$, and $W$ of each white dwarf but calculate its orbit 
in the Galactic disk. 
This allows us to define new sophisticated criteria for 
classifying thin, thick disk, and halo populations by considering orbits and 
kinematic parameters. 
Another important question is how errors of the input parameters affect 
errors of the kinematic output parameters. An error propagation code 
using a Monte Carlo simulation has been developed which allows us to check 
the statistical significance of our results. 
We use our sample to test the results of the samples which lack radial 
velocity information by investigating the effect of setting our 
radial velocities arbitrarily to zero.

Our publication is structured as follows: 
Sect.~\ref{data} deals with the data and the error analysis. 
In Sect.~\ref{method} our analysis method is described and the calibration
sample is presented.
Sect.~\ref{stud} is dedicated to   
our kinematic studies, focusing on classical and more 
sophisticated analysis methods. 
Our results appear in Sect.~\ref{results} and are discussed in 
Sect.~\ref{dis}. We finish with conclusions in Sect.~\ref{con}. 

\section{Input data and error treatment\label{data}}
\subsection{Input data \label{obs}}
\begin{figure}
  \resizebox{\hsize}{!}{
    \begin{psfrags}
      \psfrag{B}{$B$}
      \psfrag{log(N(<B))}{${\rm log}(N(<B))$}
      \includegraphics{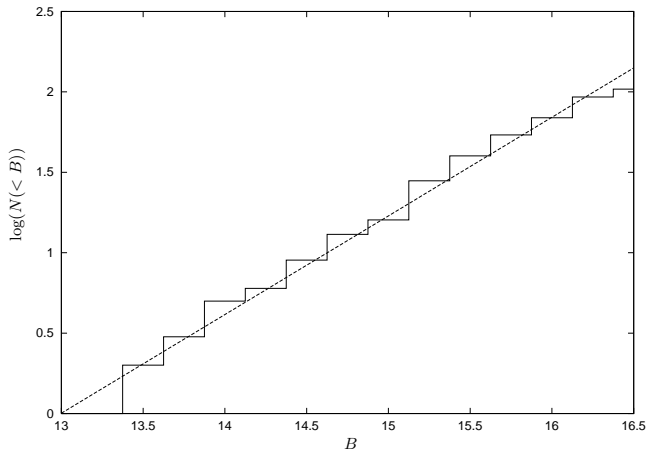}
    \end{psfrags}
    }
  \caption{Cumulative Number density as a function of $B$-magnitude}
  \label{Bhisto}
\end{figure}

Our sample consists of 107 DA white dwarfs from the SPY project 
analysed by \citet{koester01}. 
Input stars for the SPY project were drawn from the 
catalogue of \citet{mccook99}, 
the Montreal-Cambridge-Tololo (MCT) survey \citep{lamontagne00}, 
the Edinburgh-Cape (EC) survey \citep{kilkenny97}, the Hamburg/ESO
Survey \citep{wisotzki96,wisotzki00,christlieb01}, 
and the Hamburg Quasar Survey \citep{hagen95, homeier98}. 
The selection criteria were spectroscopic confirmation as white dwarf 
(at least from objective prism spectra) and 
$B < 16.5$ or $V <16.5$, respectively. 
As the stars are compiled from a number of different sources we 
cannot expect completeness. 
To make more quantitative statements on the level of incompleteness 
we compared our sample with the subsample of DA white dwarfs from the 
Palomar-Green Ultraviolet-Excess Catalog \citep{green86}. 
In Fig.~\ref{Bhisto} the cumulative number density is plotted as 
a function of $B$-magnitude. A linear function is fitted to 
the curve to obtain the count slope 
${{{\rm d}({\rm log}N)} \over {{\rm d}B}}=0.61 \pm 0.02$. 
This value does not differ much from $0.57$ derived by \citet{green86}. 
They estimated the completeness level for DA white dwarfs to be $77\%$. 
Thus, though it is difficult to compare the two samples this comparison 
demonstrates that at least 
no magnitude dependent bias is produced in the SPY sample.   

Radial velocities and spectroscopic distances were taken from 
Napiwotzki et al. (2003, in prep.). 
The aim of the SPY project is to detect radial velocity (RV) variable 
close binary white dwarfs. Two spectra at different epochs are taken 
and checked for RV variations. 
Since orbital motions distort the measurement of space motions, 
RV variable stars were discarded from our sample. 

Proper motion components $\mu_{\alpha}{\rm cos}\delta$ and 
$\mu_{\delta}$ were obtained from photographic plate data from 
the ESO Online Digitized Sky Survey (DSS1, DSS2)
\footnote{the DSS1/DSS2 data can be downloaded from 
e.g.\ {\tt http://archive.eso.org/dss/dss}} 
and the USNO catalogue \citep{monet98}. 
The Digitized Sky Survey (DSS) is a collection of (red) photographic 
Schmidt plate that have been digitised.  
One pixel corresponds to $1.7\arcsec$ on the DSS1 and to 
$1\arcsec$ on the DSS2 plates. 
In order to demonstrate that DSS data are suitable for astrometry 
we compared coordinates obtained on DSS1 plates with two different 
software packages, SExtractor \citep{bertin96} and DAOPHOT \citep{stetson92}. 
The sigma of the differences in the coordinates was $0.02\,{\rm pixel}$, 
i.e.\ $0.035\arcsec$. This indicates that our results are robust, i.e.\,they do not depend much 
on the star extraction software used. 

Each star identified on the images was assigned a number 
and $x$, $y$- coordinates were determined for the centroid of the star 
using SExtractor software \citep{bertin96}. 
From the USNO archive a catalogue of stars (with $\alpha$ and $\delta$) 
within a region of $10\arcmin$ around the coordinates of the white dwarf was extracted.
The next step was to identify some stars on the DSS1/DSS2 image with the 
corresponding USNO stars. 
After that an astrometric solution for the
plates was created and proper motions were calculated with a 
software package developed by \citet{geffert97}.
We did not measure absolute but only relative proper motions. 
To obtain absolute proper motions it would be necessary to use 
background galaxies on the plates as a reference.  
But it is difficult to identify faint galaxies on DSS images. 
So we lack the link to the extragalactic system. 
We take this into account by adopting a systematic error of 
$5\,{\rm mas\,{yr}^{-1}}$. For details we refer to \citet{altmann02}.  

To the systematic error we add an error depending on the epoch difference 
$\Delta t$ (in years) 
of the DSS1 and DSS2 plates of $200\,{\rm mas\,{yr}^{-1}}/\Delta t$. 
We assumed an accuracy of $0.1\,{\rm pixel}$ for measuring positions on 
the DSS plates (this assumption is more conservative than the 
$0.02\,{\rm pixel}$ mentioned above). 
This accuracy corresponds to $170\,{\rm mas}$ and 
$100\,{\rm mas}$ on the DSS1 and DSS2 plates, respectively. 
The total error is, by linear error propagation,
 $\sqrt{170^2+100^2}=200\,{\rm mas}$. 
We divide it by the epoch difference $\Delta t$ to obtain the 
error in $\mu_{\alpha}{\rm cos}\delta$ and $\mu_{\delta}$. 
 
For our sample this results in an error of about 
$18\,{\rm mas\,{yr}^{-1}}$ for a 
typical  epoch difference of $15\,{\rm yr}$. 
Since most DSS1 plates of the Southern hemisphere were taken in the 
seventies, the typical epoch difference does not exceed $15\,$years 
which is the limiting factor for the accuracy of the proper motion 
measurements. Nevertheless as our sample stars are near 
($d \le 500\,{\rm pc}$), 
the proper motion errors do not lead to high errors in the 
tangential velocities.  

Additional proper motions were extracted from the UCAC catalogue 
\citep{zacharias00} for stars for which no DSS2 data are available. 
The UCAC programme is an ongoing, astrometric, observational program, 
which started in February 1998. 
A global sky coverage is expected by end of 2003. 
For those stars of our sample where both UCAC and our own 
measurements of proper motions 
were available we compared the values of $\mu_{\alpha}{\rm cos}\delta$ and 
$\mu_{\delta}$ and found them to be in accordance (within the error ranges). 
This demonstrates that our error margins have been chosen sensibly and 
provides a successful test for the reliability of our astrometric measurements.  

The input parameters radial velocities, spectroscopic distances 
(Napiwotzki et al. 2003, in prep.) and 
proper motion components together with their errors are listed 
for all white dwarfs in Table~\ref{inputkin}. 
In column~2 of this table the number of spectra available for 
each white dwarf is indicated. 
If only one spectrum is available we cannot rule out a possible binarity 
of the objects. 
Nevertheless, we keep them in our sample because only about
 15\% of the white dwarfs investigated by the SPY project 
\citep{napiwotzki01} are radial velocity variable.  
From this we might expect about four stars in Table~\ref{inputkin} 
to be radial velocity (RV) variable. 
\begin{table*}[ht]
\caption[]
{Radial velocities (corrected for gravitational redshift), proper motion components and spectroscopic distances (with errors) of the white dwarfs\label{inputkin}}
\begin{tabular}{ccr@{$\pm$}lr@{$\pm$}lr@{$\pm$}lcr@{$\pm$}l}
\hline
star & spectra & \multicolumn{2}{c}{$v_{\rm rad}$} & \multicolumn{2}{r}{$\mu_{\alpha}\,{\rm cos} \delta$} & \multicolumn{2}{c}{$\mu_{\delta}$} & $d$ & \multicolumn{2}{c}{${\rm log}~d$}\\ 
  & & \multicolumn{2}{c}{${\rm km\,s^{-1}}$} & \multicolumn{2}{r}{${\rm mas\,yr^{-1}}$} & \multicolumn{2}{r}{${\rm mas\,yr^{-1}}$} & ${\rm pc}$& \multicolumn{2}{c}{${\rm pc}$} \\
\hline
HE\,0348$-$4445 & $ 2$ & \hspace*{.3cm} $ 26.5$ & $ 3.9$ & \hspace*{.8cm} $ -16$ & $  18$ & \hspace*{.45cm} $  29$ & $  18$ & $ 119.2$ & $2.08$ & $0.03$ \\
HE\,0358$-$5127 & $ 2$ & \hspace*{.3cm} $ 12.9$ & $ 3.6$ & \hspace*{.8cm} $   4$ & $  17$ & \hspace*{.45cm} $   2$ & $  17$ & $ 120.5$ & $2.08$ & $0.03$ \\
HE\,0403$-$4129 & $ 2$ & \hspace*{.3cm} $  4.7$ & $ 4.1$ & \hspace*{.8cm} $ -17$ & $  19$ & \hspace*{.45cm} $ -16$ & $  19$ & $ 146.0$ & $2.16$ & $0.03$ \\
HE\,0409$-$5154 & $ 1$ & \hspace*{.3cm} $ 46.2$ & $ 3.6$ & \hspace*{.8cm} $  42$ & $  17$ & \hspace*{.45cm} $-132$ & $  17$ & $ 147.0$ & $2.17$ & $0.03$ \\
HE\,0507$-$1855 & $ 2$ & \hspace*{.3cm} $ 17.8$ & $ 4.9$ & \hspace*{.8cm} $ -20$ & $  17$ & \hspace*{.45cm} $   0$ & $  17$ & $ 116.0$ & $2.06$ & $0.03$ \\
HE\,0532$-$5605 & $ 1$ & \hspace*{.3cm} $ 15.2$ & $ 5.2$ & \hspace*{.8cm} $  31$ & $  18$ & \hspace*{.45cm} $   1$ & $  18$ & $  52.2$ & $1.72$ & $0.03$ \\
HE\,1012$-$0049 & $ 1$ & \hspace*{.3cm} $ -3.7$ & $ 3.9$ & \hspace*{.8cm} $ -13$ & $  30$ & \hspace*{.45cm} $  26$ & $  30$ & $  99.7$ & $2.00$ & $0.02$ \\
HE\,1053$-$0914 & $ 1$ & \hspace*{.3cm} $  7.5$ & $ 3.4$ & \hspace*{.8cm} $ -14$ & $  28$ & \hspace*{.45cm} $ -15$ & $  28$ & $ 195.3$ & $2.29$ & $0.02$ \\
HE\,1117$-$0222 & $ 1$ & \hspace*{.3cm} $ 15.2$ & $ 3.2$ & \hspace*{.8cm} $-132$ & $  11$ & \hspace*{.45cm} $ -10$ & $  11$ & $  38.3$ & $1.58$ & $0.03$ \\
HE\,1124+0144 & $ 2$ & \hspace*{.3cm} $ 54.6$ & $ 3.1$ & \hspace*{.8cm} $-175$ & $  27$ & \hspace*{.45cm} $  46$ & $  27$ & $ 133.1$ & $2.12$ & $0.03$ \\
HE\,1152$-$1244 & $ 1$ & \hspace*{.3cm} $ 12.4$ & $ 2.7$ & \hspace*{.8cm} $ -19$ & $  25$ & \hspace*{.45cm} $ -47$ & $  25$ & $  88.7$ & $1.95$ & $0.02$ \\
HE\,1215+0227 & $ 2$ & \hspace*{.3cm} $  2.0$ & $12.0$ & \hspace*{.8cm} $ -19$ & $  20$ & \hspace*{.45cm} $ -13$ & $  20$ & $ 377.6$ & $2.58$ & $0.04$ \\
HE\,1225+0038 & $ 2$ & \hspace*{.3cm} $-24.0$ & $ 3.7$ & \hspace*{.8cm} $ -52$ & $  10$ & \hspace*{.45cm} $  98$ & $  10$ & $  29.0$ & $1.46$ & $0.03$ \\
HE\,1252$-$0202 & $ 2$ & \hspace*{.3cm} $ -2.1$ & $ 3.8$ & \hspace*{.8cm} $ -12$ & $  11$ & \hspace*{.45cm} $ -43$ & $  11$ & $ 124.0$ & $2.09$ & $0.02$ \\
HE\,1307$-$0059 & $ 2$ & \hspace*{.3cm} $ 27.2$ & $ 3.4$ & \hspace*{.8cm} $ -82$ & $  46$ & \hspace*{.45cm} $ -61$ & $  46$ & $  99.5$ & $2.00$ & $0.03$ \\
HE\,1310$-$0337 & $ 1$ & \hspace*{.3cm} $ 19.7$ & $ 3.7$ & \hspace*{.8cm} $ -47$ & $  20$ & \hspace*{.45cm} $  -8$ & $  20$ & $ 149.0$ & $2.17$ & $0.03$ \\
HE\,1315$-$1105 & $ 1$ & \hspace*{.3cm} $ -3.4$ & $ 3.6$ & \hspace*{.8cm} $-141$ & $  21$ & \hspace*{.45cm} $  12$ & $  21$ & $  39.7$ & $1.60$ & $0.03$ \\
HE\,1326$-$0041 & $ 2$ & \hspace*{.3cm} $  2.6$ & $ 3.7$ & \hspace*{.8cm} $  34$ & $  16$ & \hspace*{.45cm} $ -42$ & $  16$ & $ 132.5$ & $2.12$ & $0.03$ \\
HE\,1328$-$0535 & $ 2$ & \hspace*{.3cm} $  3.2$ & $ 7.3$ & \hspace*{.8cm} $ -20$ & $  20$ & \hspace*{.45cm} $ -15$ & $  20$ & $ 246.4$ & $2.39$ & $0.03$ \\
HE\,1335$-$0332 & $ 2$ & \hspace*{.3cm} $  1.6$ & $ 6.6$ & \hspace*{.8cm} $ -57$ & $  25$ & \hspace*{.45cm} $  12$ & $  25$ & $ 109.3$ & $2.04$ & $0.03$ \\
HE\,1413+0021 & $ 2$ & \hspace*{.3cm} $ -8.2$ & $ 4.1$ & \hspace*{.8cm} $  49$ & $  18$ & \hspace*{.45cm} $  19$ & $  18$ & $  70.7$ & $1.85$ & $0.03$ \\
HE\,1441$-$0047 & $ 2$ & \hspace*{.3cm} $-11.6$ & $ 4.7$ & \hspace*{.8cm} $  32$ & $  17$ & \hspace*{.45cm} $   4$ & $  17$ & $ 108.4$ & $2.04$ & $0.03$ \\
HE\,1518$-$0020 & $ 1$ & \hspace*{.3cm} $  5.9$ & $ 3.3$ & \hspace*{.8cm} $ -98$ & $  21$ & \hspace*{.45cm} $-113$ & $  21$ & $  73.2$ & $1.86$ & $0.02$ \\
HE\,1518$-$0344 & $ 2$ & \hspace*{.3cm} $ -0.7$ & $ 5.6$ & \hspace*{.8cm} $ -31$ & $  29$ & \hspace*{.45cm} $ -23$ & $  29$ & $ 204.8$ & $2.31$ & $0.03$ \\
HE\,1522$-$0410 & $ 2$ & \hspace*{.3cm} $-11.6$ & $ 4.2$ & \hspace*{.8cm} $  -7$ & $  29$ & \hspace*{.45cm} $  -7$ & $  29$ & $  66.0$ & $1.82$ & $0.03$ \\
WD\,0000$-$186 & $ 2$ & \hspace*{.3cm} $  5.2$ & $ 3.4$ & \hspace*{.8cm} $ -46$ & $  22$ & \hspace*{.45cm} $ -94$ & $  22$ & $ 111.4$ & $2.05$ & $0.05$ \\
WD\,0005$-$163 & $ 1$ & \hspace*{.3cm} $ -9.4$ & $ 3.7$ & \hspace*{.8cm} $ 169$ & $  22$ & \hspace*{.45cm} $ -56$ & $  22$ & $  91.3$ & $1.96$ & $0.03$ \\
WD\,0011+000 & $ 2$ & \hspace*{.3cm} $-12.0$ & $ 3.6$ & \hspace*{.8cm} $ 420$ & $  10$ & \hspace*{.45cm} $-182$ & $  10$ & $  36.5$ & $1.56$ & $0.03$ \\
WD\,0013$-$241 & $ 2$ & \hspace*{.3cm} $ -6.9$ & $ 3.2$ & \hspace*{.8cm} $-149$ & $  17$ & \hspace*{.45cm} $   2$ & $  17$ & $  86.8$ & $1.94$ & $0.03$ \\
WD\,0016$-$220 & $ 2$ & \hspace*{.3cm} $ -6.4$ & $ 2.8$ & \hspace*{.8cm} $ -53$ & $  16$ & \hspace*{.45cm} $ -58$ & $  16$ & $  68.3$ & $1.83$ & $0.03$ \\
WD\,0016$-$258 & $ 2$ & \hspace*{.3cm} $ 15.8$ & $ 3.7$ & \hspace*{.8cm} $ -12$ & $  17$ & \hspace*{.45cm} $ -64$ & $  17$ & $  66.8$ & $1.82$ & $0.05$ \\
WD\,0102$-$142 & $ 1$ & \hspace*{.3cm} $ -2.9$ & $ 3.5$ & \hspace*{.8cm} $  54$ & $   6$ & \hspace*{.45cm} $  37$ & $   5$ & $ 122.6$ & $2.09$ & $0.05$ \\
WD\,0106$-$358 & $ 2$ & \hspace*{.3cm} $  8.6$ & $ 4.4$ & \hspace*{.8cm} $   1$ & $  15$ & \hspace*{.45cm} $ -56$ & $  15$ & $  97.4$ & $1.99$ & $0.03$ \\
WD\,0108+143 & $ 1$ & \hspace*{.3cm} $ 13.1$ & $ 6.5$ & \hspace*{.8cm} $ 292$ & $  10$ & \hspace*{.45cm} $ -63$ & $  10$ & $  37.0$ & $1.57$ & $0.06$ \\
WD\,0110$-$139 & $ 2$ & \hspace*{.3cm} $  8.0$ & $ 3.8$ & \hspace*{.8cm} $   3$ & $   5$ & \hspace*{.45cm} $ -20$ & $   5$ & $ 129.2$ & $2.11$ & $0.05$ \\
WD\,0124$-$257 & $ 2$ & \hspace*{.3cm} $ 10.5$ & $ 3.5$ & \hspace*{.8cm} $  52$ & $  20$ & \hspace*{.45cm} $ -73$ & $  20$ & $ 129.4$ & $2.11$ & $0.05$ \\
WD\,0126+101 & $ 2$ & \hspace*{.3cm} $-17.6$ & $ 2.7$ & \hspace*{.8cm} $-146$ & $  16$ & \hspace*{.45cm} $-382$ & $  16$ & $  25.0$ & $1.40$ & $0.03$ \\
WD\,0129$-$205 & $ 2$ & \hspace*{.3cm} $ 33.6$ & $ 3.6$ & \hspace*{.8cm} $ 168$ & $  19$ & \hspace*{.45cm} $  24$ & $  21$ & $  63.7$ & $1.80$ & $0.05$ \\
WD\,0137$-$291 & $ 2$ & \hspace*{.3cm} $  5.7$ & $ 3.6$ & \hspace*{.8cm} $  53$ & $  20$ & \hspace*{.45cm} $  -5$ & $  20$ & $ 151.8$ & $2.18$ & $0.05$ \\
WD\,0138$-$236 & $ 2$ & \hspace*{.3cm} $ 26.0$ & $ 7.1$ & \hspace*{.8cm} $   5$ & $  20$ & \hspace*{.45cm} $  -1$ & $  20$ & $ 298.9$ & $2.48$ & $0.06$ \\
WD\,0140$-$392 & $ 2$ & \hspace*{.3cm} $ 25.7$ & $ 3.5$ & \hspace*{.8cm} $ 141$ & $  16$ & \hspace*{.45cm} $  68$ & $  16$ & $  56.8$ & $1.75$ & $0.03$ \\
WD\,0151+017 & $ 2$ & \hspace*{.3cm} $ 40.7$ & $ 3.0$ & \hspace*{.8cm} $ 296$ & $ 105$ & \hspace*{.45cm} $  10$ & $ 105$ & $  54.2$ & $1.73$ & $0.03$ \\
WD\,0204$-$233 & $ 1$ & \hspace*{.3cm} $ 76.3$ & $ 2.7$ & \hspace*{.8cm} $ 185$ & $  79$ & \hspace*{.45cm} $-116$ & $  79$ & $  79.1$ & $1.90$ & $0.03$ \\
WD\,0205$-$304 & $ 1$ & \hspace*{.3cm} $ 57.3$ & $ 3.1$ & \hspace*{.8cm} $ 167$ & $  17$ & \hspace*{.45cm} $ -26$ & $  17$ & $  93.6$ & $1.97$ & $0.03$ \\
WD\,0209+085 & $ 2$ & \hspace*{.3cm} $ 53.8$ & $ 5.4$ & \hspace*{.8cm} $  78$ & $  11$ & \hspace*{.45cm} $ -20$ & $  11$ & $ 101.7$ & $2.01$ & $0.06$ \\
WD\,0216+143 & $ 1$ & \hspace*{.3cm} $ -9.3$ & $ 3.4$ & \hspace*{.8cm} $-100$ & $  10$ & \hspace*{.45cm} $  42$ & $  10$ & $  89.6$ & $1.95$ & $0.03$ \\
WD\,0252$-$350 & $ 2$ & \hspace*{.3cm} $ 86.4$ & $ 2.4$ & \hspace*{.8cm} $  36$ & $  19$ & \hspace*{.45cm} $-343$ & $  19$ & $ 137.7$ & $2.14$ & $0.05$ \\
WD\,0339$-$035 & $ 1$ & \hspace*{.3cm} $ 50.9$ & $ 3.3$ & \hspace*{.8cm} $ 252$ & $  45$ & \hspace*{.45cm} $  58$ & $  45$ & $  54.9$ & $1.74$ & $0.03$ \\
WD\,0408$-$041 & $ 1$ & \hspace*{.3cm} $ -9.6$ & $ 3.5$ & \hspace*{.8cm} $  -6$ & $ 170$ & \hspace*{.45cm} $-103$ & $ 170$ & $  73.2$ & $1.86$ & $0.03$ \\
WD\,0416$-$550 & $ 2$ & \hspace*{.3cm} $ 19.6$ & $ 3.6$ & \hspace*{.8cm} $  49$ & $  17$ & \hspace*{.45cm} $   5$ & $  17$ & $ 226.1$ & $2.35$ & $0.05$ \\
WD\,0509$-$007 & $ 1$ & \hspace*{.3cm} $ 11.1$ & $ 3.1$ & \hspace*{.8cm} $ -73$ & $  10$ & \hspace*{.45cm} $  61$ & $  10$ & $ 116.9$ & $2.07$ & $0.03$ \\
WD\,0911$-$076 & $ 2$ & \hspace*{.3cm} $ 34.1$ & $ 3.2$ & \hspace*{.8cm} $ -63$ & $  30$ & \hspace*{.45cm} $   6$ & $  30$ & $ 121.5$ & $2.08$ & $0.03$ \\
WD\,0939$-$153 & $ 2$ & \hspace*{.3cm} $ 22.2$ & $ 2.7$ & \hspace*{.8cm} $ -35$ & $  23$ & \hspace*{.45cm} $  42$ & $  23$ & $  95.5$ & $1.98$ & $0.03$ \\
WD\,0951$-$155 & $ 2$ & \hspace*{.3cm} $  7.7$ & $ 3.0$ & \hspace*{.8cm} $ -88$ & $  23$ & \hspace*{.45cm} $ 115$ & $  23$ & $ 121.0$ & $2.08$ & $0.02$ \\
WD\,0956+020 & $ 2$ & \hspace*{.3cm} $ 40.1$ & $ 2.8$ & \hspace*{.8cm} $  45$ & $  20$ & \hspace*{.45cm} $   5$ & $  20$ & $  89.3$ & $1.95$ & $0.05$ \\
\hline
\end{tabular}
\end{table*}
\begin{table*}[ht]
\addtocounter{table}{-1}
\caption[]
{Radial velocities (corrected for gravitational redshift), proper motion components and spectroscopic distances (with errors) of the white dwarfs
\\continued from previous page\label{inputkin2}}
\begin{tabular}{ccr@{$\pm$}lr@{$\pm$}lr@{$\pm$}lcr@{$\pm$}l}
\hline
star & spectra & \multicolumn{2}{c}{$v_{\rm rad}$} & \multicolumn{2}{r}{$\mu_{\alpha}\,{\rm cos} \delta$} & \multicolumn{2}{c}{$\mu_{\delta}$} & $d$ & \multicolumn{2}{c}{${\rm log}~d$}\\ 
  & & \multicolumn{2}{c}{${\rm km\,s^{-1}}$} & \multicolumn{2}{r}{${\rm mas\,yr^{-1}}$} & \multicolumn{2}{r}{${\rm mas\,yr^{-1}}$} & ${\rm pc}$& \multicolumn{2}{c}{${\rm pc}$} \\
\hline
WD\,1015$-$216 & $ 1$ & \hspace*{.3cm} $-12.6$ & $ 5.4$ & \hspace*{.8cm} $ -13$ & $  18$ & \hspace*{.45cm} $ -21$ & $  18$ & $ 173.5$ & $2.24$ & $0.03$ \\
WD\,1020$-$207 & $ 1$ & \hspace*{.3cm} $ 56.3$ & $ 2.9$ & \hspace*{.8cm} $ -74$ & $  12$ & \hspace*{.45cm} $ -15$ & $  12$ & $  83.2$ & $1.92$ & $0.03$ \\
WD\,1102$-$183 & $ 1$ & \hspace*{.3cm} $ 24.0$ & $ 3.6$ & \hspace*{.8cm} $ 135$ & $  19$ & \hspace*{.45cm} $ -73$ & $  19$ & $  39.9$ & $1.60$ & $0.03$ \\
WD\,1122$-$324 & $ 1$ & \hspace*{.3cm} $-19.9$ & $ 3.2$ & \hspace*{.8cm} $ -84$ & $  19$ & \hspace*{.45cm} $   4$ & $  19$ & $ 127.6$ & $2.11$ & $0.03$ \\
WD\,1124$-$293 & $ 1$ & \hspace*{.3cm} $ -2.6$ & $ 3.5$ & \hspace*{.8cm} $ 225$ & $  23$ & \hspace*{.45cm} $-286$ & $  23$ & $  33.9$ & $1.53$ & $0.03$ \\
WD\,1126$-$222 & $ 1$ & \hspace*{.3cm} $ 32.5$ & $ 3.4$ & \hspace*{.8cm} $-151$ & $  18$ & \hspace*{.45cm} $ -52$ & $  18$ & $  93.3$ & $1.97$ & $0.03$ \\
WD\,1144$-$246 & $ 1$ & \hspace*{.3cm} $-14.5$ & $ 3.9$ & \hspace*{.8cm} $ -30$ & $  19$ & \hspace*{.45cm} $ -14$ & $  19$ & $ 270.7$ & $2.43$ & $0.03$ \\
WD\,1150$-$153 & $ 1$ & \hspace*{.3cm} $  1.7$ & $ 3.4$ & \hspace*{.8cm} $ -17$ & $  25$ & \hspace*{.45cm} $ -65$ & $  25$ & $  84.6$ & $1.93$ & $0.03$ \\
WD\,1155$-$243 & $ 2$ & \hspace*{.3cm} $ 22.9$ & $ 3.2$ & \hspace*{.8cm} $ -75$ & $  18$ & \hspace*{.45cm} $-107$ & $  18$ & $ 116.2$ & $2.07$ & $0.03$ \\
WD\,1159$-$098 & $ 2$ & \hspace*{.3cm} $ -7.3$ & $ 6.3$ & \hspace*{.8cm} $-156$ & $  46$ & \hspace*{.45cm} $-112$ & $  46$ & $  33.0$ & $1.52$ & $0.03$ \\
WD\,1201$-$001 & $ 2$ & \hspace*{.3cm} $ 10.8$ & $ 4.4$ & \hspace*{.8cm} $-153$ & $  54$ & \hspace*{.45cm} $  29$ & $  54$ & $  63.7$ & $1.80$ & $0.03$ \\
WD\,1204$-$322 & $ 1$ & \hspace*{.3cm} $-12.0$ & $ 3.4$ & \hspace*{.8cm} $ -70$ & $  11$ & \hspace*{.45cm} $ -31$ & $  11$ & $ 103.0$ & $2.01$ & $0.03$ \\
WD\,1207$-$157 & $ 2$ & \hspace*{.3cm} $-14.2$ & $ 3.1$ & \hspace*{.8cm} $ -60$ & $  25$ & \hspace*{.45cm} $ -18$ & $  25$ & $ 141.3$ & $2.15$ & $0.02$ \\
WD\,1244$-$125 & $ 1$ & \hspace*{.3cm} $  2.7$ & $ 3.0$ & \hspace*{.8cm} $-180$ & $  25$ & \hspace*{.45cm} $  -3$ & $  25$ & $  48.6$ & $1.69$ & $0.03$ \\
WD\,1326$-$236 & $ 2$ & \hspace*{.3cm} $-16.7$ & $ 3.3$ & \hspace*{.8cm} $ -49$ & $  17$ & \hspace*{.45cm} $ -14$ & $  17$ & $  93.6$ & $1.97$ & $0.03$ \\
WD\,1334$-$678 & $ 1$ & \hspace*{.3cm} $ 28.9$ & $ 3.9$ & \hspace*{.8cm} $-607$ & $  56$ & \hspace*{.45cm} $ -18$ & $  56$ & $  37.1$ & $1.57$ & $0.03$ \\
WD\,1342$-$237 & $ 2$ & \hspace*{.3cm} $ 15.3$ & $ 4.1$ & \hspace*{.8cm} $ -20$ & $  17$ & \hspace*{.45cm} $  -2$ & $  17$ & $  64.7$ & $1.81$ & $0.03$ \\
WD\,1344+106 & $ 1$ & \hspace*{.3cm} $-60.2$ & $ 4.1$ & \hspace*{.8cm} $-886$ & $  10$ & \hspace*{.45cm} $-144$ & $  10$ & $  18.2$ & $1.26$ & $0.03$ \\
WD\,1356$-$233 & $ 1$ & \hspace*{.3cm} $-57.4$ & $ 3.8$ & \hspace*{.8cm} $-324$ & $  19$ & \hspace*{.45cm} $ -50$ & $  19$ & $  31.2$ & $1.49$ & $0.03$ \\
WD\,1401$-$147 & $ 2$ & \hspace*{.3cm} $-13.6$ & $ 3.5$ & \hspace*{.8cm} $-160$ & $  21$ & \hspace*{.45cm} $-116$ & $  21$ & $  59.3$ & $1.77$ & $0.03$ \\
WD\,1422+095 & $ 2$ & \hspace*{.3cm} $-29.2$ & $ 3.3$ & \hspace*{.8cm} $-211$ & $  10$ & \hspace*{.45cm} $-147$ & $  10$ & $  36.9$ & $1.57$ & $0.03$ \\
WD\,1426$-$276 & $ 2$ & \hspace*{.3cm} $ 56.8$ & $ 2.6$ & \hspace*{.8cm} $  -3$ & $  23$ & \hspace*{.45cm} $-228$ & $  23$ & $ 123.4$ & $2.09$ & $0.03$ \\
WD\,1448+077 & $ 1$ & \hspace*{.3cm} $-118.5$ & $ 2.7$ & \hspace*{.8cm} $-816$ & $  11$ & \hspace*{.45cm} $-454$ & $  11$ & $  85.0$ & $1.93$ & $0.03$ \\
WD\,1457$-$086 & $ 2$ & \hspace*{.3cm} $ -6.1$ & $ 3.5$ & \hspace*{.8cm} $  11$ & $  20$ & \hspace*{.45cm} $ -42$ & $  20$ & $ 118.6$ & $2.07$ & $0.03$ \\
WD\,1515$-$164 & $ 1$ & \hspace*{.3cm} $ 23.8$ & $ 3.1$ & \hspace*{.8cm} $ 105$ & $  27$ & \hspace*{.45cm} $ -10$ & $  27$ & $  97.1$ & $1.99$ & $0.03$ \\
WD\,1524$-$749 & $ 2$ & \hspace*{.3cm} $187.2$ & $ 3.0$ & \hspace*{.8cm} $-407$ & $  16$ & \hspace*{.45cm} $-254$ & $  16$ & $ 165.5$ & $2.22$ & $0.03$ \\
WD\,1537$-$152 & $ 2$ & \hspace*{.3cm} $-22.6$ & $ 3.3$ & \hspace*{.8cm} $ -84$ & $  38$ & \hspace*{.45cm} $  17$ & $  38$ & $  98.5$ & $1.99$ & $0.02$ \\
WD\,1555$-$089 & $ 2$ & \hspace*{.3cm} $ 44.0$ & $ 3.9$ & \hspace*{.8cm} $ -66$ & $  25$ & \hspace*{.45cm} $-145$ & $  25$ & $  58.5$ & $1.77$ & $0.03$ \\
WD\,1609+135 & $ 1$ & \hspace*{.3cm} $ 26.4$ & $ 6.1$ & \hspace*{.8cm} $  17$ & $  10$ & \hspace*{.45cm} $-537$ & $  10$ & $  21.6$ & $1.33$ & $0.03$ \\
WD\,1636+057 & $ 2$ & \hspace*{.3cm} $ -8.4$ & $ 6.9$ & \hspace*{.8cm} $-307$ & $  10$ & \hspace*{.45cm} $-418$ & $  10$ & $  24.9$ & $1.40$ & $0.03$ \\
WD\,1834$-$781 & $ 2$ & \hspace*{.3cm} $ 71.5$ & $ 2.8$ & \hspace*{.8cm} $ 130$ & $  16$ & \hspace*{.45cm} $-313$ & $  16$ & $  92.8$ & $1.97$ & $0.03$ \\
WD\,1952$-$206 & $ 2$ & \hspace*{.3cm} $ 18.0$ & $ 2.9$ & \hspace*{.8cm} $ 109$ & $  19$ & \hspace*{.45cm} $-380$ & $  19$ & $  60.7$ & $1.78$ & $0.03$ \\
WD\,1959+059 & $ 2$ & \hspace*{.3cm} $-40.9$ & $ 4.5$ & \hspace*{.8cm} $-120$ & $  10$ & \hspace*{.45cm} $-110$ & $  10$ & $  80.4$ & $1.91$ & $0.03$ \\
WD\,2007$-$219 & $ 2$ & \hspace*{.3cm} $-55.9$ & $ 3.6$ & \hspace*{.8cm} $ 108$ & $  19$ & \hspace*{.45cm} $-308$ & $  19$ & $  25.5$ & $1.41$ & $0.03$ \\
WD\,2014$-$575 & $ 1$ & \hspace*{.3cm} $ 40.9$ & $ 3.4$ & \hspace*{.8cm} $  69$ & $  19$ & \hspace*{.45cm} $ -39$ & $  19$ & $  44.5$ & $1.65$ & $0.03$ \\
WD\,2139+115 & $ 1$ & \hspace*{.3cm} $ -0.6$ & $ 3.1$ & \hspace*{.8cm} $ 204$ & $  10$ & \hspace*{.45cm} $ -25$ & $  10$ & $  98.9$ & $2.00$ & $0.03$ \\
WD\,2151$-$307 & $ 2$ & \hspace*{.3cm} $  6.0$ & $ 5.0$ & \hspace*{.8cm} $  37$ & $  17$ & \hspace*{.45cm} $ -12$ & $  17$ & $  85.7$ & $1.93$ & $0.03$ \\
WD\,2254+126 & $ 2$ & \hspace*{.3cm} $-10.6$ & $ 4.2$ & \hspace*{.8cm} $ 186$ & $  10$ & \hspace*{.45cm} $   1$ & $  10$ & $  63.0$ & $1.80$ & $0.05$ \\
WD\,2306+124 & $ 2$ & \hspace*{.3cm} $ 13.8$ & $ 3.6$ & \hspace*{.8cm} $  -7$ & $  10$ & \hspace*{.45cm} $  23$ & $  10$ & $  85.4$ & $1.93$ & $0.03$ \\
WD\,2311$-$260 & $ 2$ & \hspace*{.3cm} $ 25.7$ & $ 6.8$ & \hspace*{.8cm} $  12$ & $  18$ & \hspace*{.45cm} $  -7$ & $  18$ & $ 434.9$ & $2.64$ & $0.04$ \\
WD\,2318$-$226 & $ 2$ & \hspace*{.3cm} $ 15.5$ & $ 5.0$ & \hspace*{.8cm} $   6$ & $  18$ & \hspace*{.45cm} $ -51$ & $  18$ & $ 209.7$ & $2.32$ & $0.03$ \\
WD\,2322$-$181 & $ 2$ & \hspace*{.3cm} $ -0.3$ & $ 3.2$ & \hspace*{.8cm} $ 242$ & $  19$ & \hspace*{.45cm} $  10$ & $  19$ & $ 101.0$ & $2.00$ & $0.05$ \\
WD\,2326+049 & $ 1$ & \hspace*{.3cm} $ 15.3$ & $ 3.0$ & \hspace*{.8cm} $-412$ & $  10$ & \hspace*{.45cm} $-263$ & $  10$ & $  19.2$ & $1.28$ & $0.03$ \\
WD\,2329$-$332 & $ 2$ & \hspace*{.3cm} $ 13.1$ & $ 3.7$ & \hspace*{.8cm} $  19$ & $  22$ & \hspace*{.45cm} $  37$ & $  22$ & $ 157.5$ & $2.20$ & $0.05$ \\
WD\,2333$-$049 & $ 2$ & \hspace*{.3cm} $ 11.3$ & $ 3.7$ & \hspace*{.8cm} $-193$ & $  37$ & \hspace*{.45cm} $-124$ & $  37$ & $  53.5$ & $1.73$ & $0.03$ \\
WD\,2333$-$165 & $ 1$ & \hspace*{.3cm} $ 47.5$ & $ 2.6$ & \hspace*{.8cm} $  87$ & $  30$ & \hspace*{.45cm} $-147$ & $  30$ & $  31.8$ & $1.50$ & $0.03$ \\
WD\,2347$-$192 & $ 1$ & \hspace*{.3cm} $ 12.9$ & $ 4.1$ & \hspace*{.8cm} $ -46$ & $  19$ & \hspace*{.45cm} $  -8$ & $  19$ & $ 167.9$ & $2.23$ & $0.03$ \\
WD\,2348$-$244 & $ 2$ & \hspace*{.3cm} $ 36.5$ & $ 3.5$ & \hspace*{.8cm} $ -94$ & $  19$ & \hspace*{.45cm} $-237$ & $  19$ & $  49.7$ & $1.70$ & $0.03$ \\
WD\,2349$-$283 & $ 2$ & \hspace*{.3cm} $ 17.1$ & $ 2.9$ & \hspace*{.8cm} $  44$ & $  17$ & \hspace*{.45cm} $ -62$ & $  17$ & $  96.7$ & $1.99$ & $0.05$ \\
WD\,2350$-$248 & $ 2$ & \hspace*{.3cm} $ 10.8$ & $ 5.5$ & \hspace*{.8cm} $ -13$ & $  19$ & \hspace*{.45cm} $   8$ & $  19$ & $  99.9$ & $2.00$ & $0.05$ \\
WD\,2351$-$368 & $ 2$ & \hspace*{.3cm} $-26.9$ & $ 3.0$ & \hspace*{.8cm} $  33$ & $  17$ & \hspace*{.45cm} $-680$ & $  17$ & $  62.6$ & $1.80$ & $0.05$ \\
WD\,2354$-$151 & $ 1$ & \hspace*{.3cm} $  7.3$ & $ 3.6$ & \hspace*{.8cm} $  -1$ & $  22$ & \hspace*{.45cm} $ -23$ & $  22$ & $ 278.9$ & $2.45$ & $0.05$ \\
\hline
\end{tabular}
\end{table*}

\begin{table*}
\caption[]
{Kinematic parameters of the white dwarfs\label{kinpar}}
\begin{tabular}{cr@{$\pm$}lr@{$\pm$}lr@{$\pm$}lr@{$\pm$}lr@{$\pm$}l}
\hline
star & \multicolumn{2}{c}{$e$} & \multicolumn{2}{c}{$J_z$} & \multicolumn{2}{c}{$U$} & \multicolumn{2}{c}{$V$} & \multicolumn{2}{c}{$W$} \\ 
 & \multicolumn{2}{c}{}& \multicolumn{2}{c}{${\rm kpc\,km\,s^{-1}}$} & \multicolumn{2}{c}{${\rm km\,s^{-1}}$} & \multicolumn{2}{c}{${\rm km\,s^{-1}}$} & \multicolumn{2}{c}{${\rm km\,s^{-1}}$} \\
\hline
HE\,0348$-$4445 & $0.04$ & $0.02$ & \hspace*{.4cm} $ 1898$ & $   59$ & $-4.6$ & $ 8.0$ & $222.6$ & $ 6.9$ & $-20.9$ & $ 6.1$ \\
HE\,0358$-$5127 & $0.05$ & $0.02$ & \hspace*{.4cm} $ 1837$ & $   53$ & $ 7.4$ & $ 7.8$ & $215.8$ & $ 6.2$ & $-1.5$ & $ 5.6$ \\
HE\,0403$-$4129 & $0.10$ & $0.03$ & \hspace*{.4cm} $ 1922$ & $   75$ & $24.6$ & $10.6$ & $225.3$ & $ 8.8$ & $-4.5$ & $ 7.5$ \\
HE\,0409$-$5154 & $0.43$ & $0.03$ & \hspace*{.4cm} $ 1220$ & $   63$ & $81.3$ & $10.4$ & $144.3$ & $ 7.4$ & $ 9.4$ & $ 7.1$ \\
HE\,0507$-$1855 & $0.03$ & $0.02$ & \hspace*{.4cm} $ 1903$ & $   60$ & $ 0.8$ & $ 6.4$ & $221.9$ & $ 6.9$ & $-10.9$ & $ 6.9$ \\
HE\,0532$-$5605 & $0.06$ & $0.02$ & \hspace*{.4cm} $ 1774$ & $   35$ & $ 9.1$ & $ 3.6$ & $208.7$ & $ 4.2$ & $ 5.3$ & $ 3.9$ \\
HE\,1012$-$0049 & $0.08$ & $0.03$ & \hspace*{.4cm} $ 2009$ & $   75$ & $ 1.3$ & $10.6$ & $235.4$ & $ 8.8$ & $ 6.7$ & $ 8.6$ \\
HE\,1053$-$0914 & $0.10$ & $0.05$ & \hspace*{.4cm} $ 1775$ & $  123$ & $ 5.7$ & $20.0$ & $208.4$ & $14.4$ & $-0.5$ & $14.9$ \\
HE\,1117$-$0222 & $0.07$ & $0.01$ & \hspace*{.4cm} $ 1756$ & $   18$ & $-10.1$ & $ 1.9$ & $206.5$ & $ 2.1$ & $ 9.9$ & $ 2.4$ \\
HE\,1124+0144 & $0.37$ & $0.04$ & \hspace*{.4cm} $ 1519$ & $  102$ & $-103.8$ & $15.1$ & $177.6$ & $12.0$ & $30.3$ & $ 7.9$ \\
HE\,1152$-$1244 & $0.10$ & $0.03$ & \hspace*{.4cm} $ 1711$ & $   57$ & $14.4$ & $ 8.2$ & $201.7$ & $ 6.7$ & $ 1.9$ & $ 6.1$ \\
HE\,1215+0227 & $0.18$ & $0.09$ & \hspace*{.4cm} $ 1601$ & $  227$ & $-7.8$ & $28.8$ & $189.1$ & $26.8$ & $-5.2$ & $15.6$ \\
HE\,1225+0038 & $0.10$ & $0.01$ & \hspace*{.4cm} $ 2057$ & $   14$ & $-5.5$ & $ 1.4$ & $242.1$ & $ 1.7$ & $-8.9$ & $ 2.7$ \\
HE\,1252$-$0202 & $0.10$ & $0.02$ & \hspace*{.4cm} $ 1722$ & $   43$ & $16.5$ & $ 5.2$ & $203.5$ & $ 5.0$ & $-7.0$ & $ 3.8$ \\
HE\,1307$-$0059 & $0.22$ & $0.07$ & \hspace*{.4cm} $ 1454$ & $  140$ & $ 1.6$ & $16.8$ & $171.7$ & $16.5$ & $19.7$ & $ 9.1$ \\
HE\,1310$-$0337 & $0.12$ & $0.05$ & \hspace*{.4cm} $ 1643$ & $   94$ & $-7.0$ & $11.3$ & $194.4$ & $11.1$ & $23.8$ & $ 6.6$ \\
HE\,1315$-$1105 & $0.05$ & $0.01$ & \hspace*{.4cm} $ 1808$ & $   29$ & $-12.9$ & $ 3.5$ & $213.0$ & $ 3.4$ & $ 8.5$ & $ 3.1$ \\
HE\,1326$-$0041 & $0.14$ & $0.02$ & \hspace*{.4cm} $ 1834$ & $   65$ & $42.0$ & $ 8.0$ & $217.3$ & $ 7.7$ & $-5.7$ & $ 4.7$ \\
HE\,1328$-$0535 & $0.13$ & $0.06$ & \hspace*{.4cm} $ 1658$ & $  155$ & $ 3.0$ & $17.8$ & $197.7$ & $18.5$ & $ 3.0$ & $11.7$ \\
HE\,1335$-$0332 & $0.07$ & $0.03$ & \hspace*{.4cm} $ 1785$ & $   85$ & $-14.1$ & $ 9.9$ & $211.1$ & $10.0$ & $16.9$ & $ 7.5$ \\
HE\,1413+0021 & $0.11$ & $0.02$ & \hspace*{.4cm} $ 2047$ & $   43$ & $15.3$ & $ 4.7$ & $242.0$ & $ 5.1$ & $-2.4$ & $ 4.0$ \\
HE\,1441$-$0047 & $0.10$ & $0.03$ & \hspace*{.4cm} $ 2015$ & $   61$ & $13.3$ & $ 6.2$ & $239.0$ & $ 7.3$ & $-8.2$ & $ 5.2$ \\
HE\,1518$-$0020 & $0.22$ & $0.03$ & \hspace*{.4cm} $ 1467$ & $   52$ & $16.7$ & $ 4.7$ & $173.6$ & $ 6.1$ & $12.3$ & $ 4.6$ \\
HE\,1518$-$0344 & $0.16$ & $0.09$ & \hspace*{.4cm} $ 1578$ & $  191$ & $ 5.3$ & $15.7$ & $189.0$ & $22.9$ & $11.1$ & $17.3$ \\
HE\,1522$-$0410 & $0.03$ & $0.02$ & \hspace*{.4cm} $ 1878$ & $   64$ & $ 2.0$ & $ 5.4$ & $222.2$ & $ 7.6$ & $-0.3$ & $ 6.0$ \\
WD\,0000$-$186 & $0.21$ & $0.04$ & \hspace*{.4cm} $ 1650$ & $   82$ & $56.7$ & $10.3$ & $194.2$ & $ 9.7$ & $-1.4$ & $ 3.7$ \\
WD\,0005$-$163 & $0.27$ & $0.03$ & \hspace*{.4cm} $ 1432$ & $   66$ & $-41.4$ & $ 7.8$ & $168.7$ & $ 7.8$ & $-1.0$ & $ 3.5$ \\
WD\,0011+000 & $0.29$ & $0.02$ & \hspace*{.4cm} $ 1376$ & $   30$ & $-35.5$ & $ 2.8$ & $161.9$ & $ 3.6$ & $-7.7$ & $ 3.1$ \\
WD\,0013$-$241 & $0.26$ & $0.02$ & \hspace*{.4cm} $ 2161$ & $   51$ & $63.5$ & $ 6.2$ & $254.4$ & $ 6.0$ & $22.5$ & $ 2.7$ \\
WD\,0016$-$220 & $0.11$ & $0.01$ & \hspace*{.4cm} $ 1842$ & $   34$ & $34.8$ & $ 4.1$ & $216.7$ & $ 4.0$ & $13.7$ & $ 2.4$ \\
WD\,0016$-$258 & $0.09$ & $0.02$ & \hspace*{.4cm} $ 1791$ & $   40$ & $25.8$ & $ 4.6$ & $210.9$ & $ 4.7$ & $-8.9$ & $ 3.1$ \\
WD\,0102$-$142 & $0.09$ & $0.01$ & \hspace*{.4cm} $ 1913$ & $   20$ & $-26.5$ & $ 4.6$ & $224.6$ & $ 2.4$ & $16.5$ & $ 2.8$ \\
WD\,0106$-$358 & $0.12$ & $0.02$ & \hspace*{.4cm} $ 1722$ & $   49$ & $25.6$ & $ 5.6$ & $202.7$ & $ 5.8$ & $ 2.5$ & $ 3.6$ \\
WD\,0108+143 & $0.15$ & $0.02$ & \hspace*{.4cm} $ 1661$ & $   41$ & $-32.3$ & $ 4.8$ & $195.1$ & $ 4.9$ & $-6.1$ & $ 4.1$ \\
WD\,0110$-$139 & $0.05$ & $0.01$ & \hspace*{.4cm} $ 1838$ & $   23$ & $14.5$ & $ 2.6$ & $215.5$ & $ 2.8$ & $-3.3$ & $ 3.1$ \\
WD\,0124$-$257 & $0.23$ & $0.05$ & \hspace*{.4cm} $ 1443$ & $  100$ & $11.0$ & $10.3$ & $169.5$ & $11.7$ & $-0.5$ & $ 3.2$ \\
WD\,0126+101 & $0.19$ & $0.01$ & \hspace*{.4cm} $ 1680$ & $   17$ & $51.3$ & $ 2.4$ & $197.3$ & $ 2.0$ & $-9.4$ & $ 2.6$ \\
WD\,0129$-$205 & $0.15$ & $0.02$ & \hspace*{.4cm} $ 1707$ & $   45$ & $-39.8$ & $ 6.3$ & $200.5$ & $ 5.3$ & $-17.2$ & $ 3.1$ \\
WD\,0137$-$291 & $0.12$ & $0.05$ & \hspace*{.4cm} $ 1683$ & $  105$ & $-17.7$ & $12.5$ & $197.5$ & $12.4$ & $ 8.9$ & $ 3.9$ \\
WD\,0138$-$236 & $0.12$ & $0.07$ & \hspace*{.4cm} $ 1870$ & $  204$ & $ 1.4$ & $24.6$ & $218.5$ & $23.9$ & $-17.3$ & $ 7.6$ \\
WD\,0140$-$392 & $0.11$ & $0.01$ & \hspace*{.4cm} $ 1773$ & $   29$ & $-30.4$ & $ 4.1$ & $208.5$ & $ 3.4$ & $-13.7$ & $ 2.9$ \\
WD\,0151+017 & $0.26$ & $0.07$ & \hspace*{.4cm} $ 1613$ & $  183$ & $-66.1$ & $18.8$ & $189.3$ & $21.4$ & $-7.8$ & $11.8$ \\
WD\,0204$-$233 & $0.38$ & $0.11$ & \hspace*{.4cm} $ 1178$ & $  207$ & $-30.7$ & $22.6$ & $138.2$ & $24.3$ & $-49.8$ & $ 7.4$ \\
WD\,0205$-$304 & $0.31$ & $0.03$ & \hspace*{.4cm} $ 1341$ & $   53$ & $-45.1$ & $ 6.0$ & $157.3$ & $ 6.3$ & $-26.1$ & $ 3.0$ \\
WD\,0209+085 & $0.15$ & $0.02$ & \hspace*{.4cm} $ 1787$ & $   48$ & $-43.6$ & $ 5.0$ & $209.0$ & $ 5.7$ & $-27.7$ & $ 4.3$ \\
WD\,0216+143 & $0.23$ & $0.02$ & \hspace*{.4cm} $ 2242$ & $   34$ & $40.2$ & $ 3.3$ & $261.8$ & $ 3.9$ & $10.9$ & $ 3.1$ \\
WD\,0252$-$350 & $0.90$ & $0.06$ & \hspace*{.4cm} $  268$ & $  161$ & $146.6$ & $18.5$ & $32.4$ & $18.8$ & $-45.1$ & $ 5.6$ \\
WD\,0339$-$035 & $0.25$ & $0.03$ & \hspace*{.4cm} $ 1592$ & $   84$ & $-63.5$ & $ 7.2$ & $186.4$ & $ 9.8$ & $18.8$ & $ 7.7$ \\
WD\,0408$-$041 & $0.24$ & $0.12$ & \hspace*{.4cm} $ 1751$ & $  393$ & $37.7$ & $30.8$ & $204.7$ & $46.0$ & $-3.4$ & $38.0$ \\
WD\,0416$-$550 & $0.19$ & $0.05$ & \hspace*{.4cm} $ 1526$ & $   98$ & $-13.1$ & $14.5$ & $179.0$ & $11.5$ & $29.2$ & $12.0$ \\
WD\,0509$-$007 & $0.23$ & $0.02$ & \hspace*{.4cm} $ 2326$ & $   45$ & $-9.4$ & $ 3.3$ & $270.3$ & $ 5.1$ & $-14.3$ & $ 4.7$ \\
WD\,0911$-$076 & $0.13$ & $0.03$ & \hspace*{.4cm} $ 1727$ & $   78$ & $-31.9$ & $12.5$ & $201.5$ & $ 9.1$ & $-1.9$ & $12.7$ \\
WD\,0939$-$153 & $0.06$ & $0.02$ & \hspace*{.4cm} $ 1839$ & $   43$ & $-18.3$ & $ 8.0$ & $215.5$ & $ 5.0$ & $18.0$ & $ 7.8$ \\
WD\,0951$-$155 & $0.25$ & $0.04$ & \hspace*{.4cm} $ 2124$ & $   58$ & $-67.4$ & $11.3$ & $248.2$ & $ 6.7$ & $19.4$ & $ 9.6$ \\
WD\,0956+020 & $0.07$ & $0.03$ & \hspace*{.4cm} $ 1744$ & $   52$ & $ 8.6$ & $ 6.7$ & $204.3$ & $ 6.0$ & $46.1$ & $ 5.7$ \\
\hline
\end{tabular}
\end{table*}
\begin{table*}
\addtocounter{table}{-1}
\caption[]
{Kinematic parameters of the white dwarfs\\
continued from previous page\label{kinpar2}}
\begin{tabular}{cr@{$\pm$}lr@{$\pm$}lr@{$\pm$}lr@{$\pm$}lr@{$\pm$}l}
\hline
star & \multicolumn{2}{c}{$e$} & \multicolumn{2}{c}{$J_z$} & \multicolumn{2}{c}{$U$} & \multicolumn{2}{c}{$V$} & \multicolumn{2}{c}{$W$} \\ 
 & \multicolumn{2}{c}{}& \multicolumn{2}{c}{${\rm kpc\,km\,s^{-1}}$} & \multicolumn{2}{c}{${\rm km\,s^{-1}}$} & \multicolumn{2}{c}{${\rm km\,s^{-1}}$} & \multicolumn{2}{c}{${\rm km\,s^{-1}}$} \\
\hline
WD\,1015$-$216 & $0.08$ & $0.03$ & \hspace*{.4cm} $ 1928$ & $   64$ & $14.1$ & $12.6$ & $226.5$ & $ 7.5$ & $-16.4$ & $11.2$ \\
WD\,1020$-$207 & $0.24$ & $0.01$ & \hspace*{.4cm} $ 1432$ & $   25$ & $-16.2$ & $ 3.9$ & $168.2$ & $ 2.9$ & $15.4$ & $ 3.7$ \\
WD\,1102$-$183 & $0.14$ & $0.01$ & \hspace*{.4cm} $ 1760$ & $   25$ & $40.0$ & $ 3.5$ & $207.2$ & $ 3.0$ & $22.5$ & $ 3.0$ \\
WD\,1122$-$324 & $0.12$ & $0.03$ & \hspace*{.4cm} $ 1909$ & $   44$ & $-38.1$ & $ 9.5$ & $224.7$ & $ 5.2$ & $-17.4$ & $ 8.7$ \\
WD\,1124$-$293 & $0.21$ & $0.01$ & \hspace*{.4cm} $ 1897$ & $   24$ & $63.0$ & $ 4.2$ & $223.5$ & $ 2.8$ & $-19.9$ & $ 3.2$ \\
WD\,1126$-$222 & $0.27$ & $0.02$ & \hspace*{.4cm} $ 1391$ & $   40$ & $-32.4$ & $ 7.0$ & $163.6$ & $ 4.8$ & $-11.6$ & $ 5.5$ \\
WD\,1144$-$246 & $0.09$ & $0.05$ & \hspace*{.4cm} $ 1795$ & $  108$ & $-18.5$ & $19.9$ & $212.1$ & $12.9$ & $-24.9$ & $16.7$ \\
WD\,1150$-$153 & $0.09$ & $0.02$ & \hspace*{.4cm} $ 1747$ & $   52$ & $17.1$ & $ 8.2$ & $205.9$ & $ 6.1$ & $-10.6$ & $ 5.9$ \\
WD\,1155$-$243 & $0.26$ & $0.03$ & \hspace*{.4cm} $ 1368$ & $   52$ & $ 5.3$ & $ 7.8$ & $161.6$ & $ 6.1$ & $-33.7$ & $ 7.4$ \\
WD\,1159$-$098 & $0.06$ & $0.02$ & \hspace*{.4cm} $ 1761$ & $   48$ & $-3.3$ & $ 5.9$ & $207.3$ & $ 5.7$ & $-13.6$ & $ 5.4$ \\
WD\,1201$-$001 & $0.13$ & $0.04$ & \hspace*{.4cm} $ 1749$ & $  100$ & $-33.5$ & $12.9$ & $205.7$ & $11.7$ & $12.0$ & $ 7.1$ \\
WD\,1204$-$322 & $0.06$ & $0.02$ & \hspace*{.4cm} $ 1807$ & $   31$ & $-17.4$ & $ 4.5$ & $213.3$ & $ 3.7$ & $-17.9$ & $ 4.3$ \\
WD\,1207$-$157 & $0.09$ & $0.04$ & \hspace*{.4cm} $ 1767$ & $   89$ & $-22.4$ & $13.1$ & $208.4$ & $10.5$ & $-18.1$ & $ 9.8$ \\
WD\,1244$-$125 & $0.11$ & $0.02$ & \hspace*{.4cm} $ 1706$ & $   36$ & $-23.0$ & $ 4.9$ & $201.1$ & $ 4.2$ & $ 8.2$ & $ 3.6$ \\
WD\,1326$-$236 & $0.05$ & $0.02$ & \hspace*{.4cm} $ 1843$ & $   46$ & $-13.6$ & $ 5.6$ & $218.0$ & $ 5.5$ & $-4.9$ & $ 5.1$ \\
WD\,1334$-$678 & $0.41$ & $0.04$ & \hspace*{.4cm} $ 1163$ & $   56$ & $-54.6$ & $ 8.2$ & $137.0$ & $ 6.6$ & $20.1$ & $ 8.0$ \\
WD\,1342$-$237 & $0.06$ & $0.01$ & \hspace*{.4cm} $ 1802$ & $   33$ & $15.7$ & $ 4.0$ & $213.0$ & $ 3.9$ & $17.1$ & $ 3.8$ \\
WD\,1344+106 & $0.28$ & $0.02$ & \hspace*{.4cm} $ 1480$ & $   26$ & $-60.8$ & $ 2.9$ & $174.2$ & $ 3.0$ & $-36.1$ & $ 3.0$ \\
WD\,1356$-$233 & $0.19$ & $0.01$ & \hspace*{.4cm} $ 1855$ & $   26$ & $-58.5$ & $ 3.1$ & $218.6$ & $ 3.0$ & $-20.5$ & $ 2.6$ \\
WD\,1401$-$147 & $0.19$ & $0.02$ & \hspace*{.4cm} $ 1524$ & $   44$ & $-18.7$ & $ 4.2$ & $180.0$ & $ 5.1$ & $-11.5$ & $ 4.1$ \\
WD\,1422+095 & $0.18$ & $0.01$ & \hspace*{.4cm} $ 1545$ & $   22$ & $-13.2$ & $ 1.8$ & $182.1$ & $ 2.5$ & $-13.9$ & $ 2.5$ \\
WD\,1426$-$276 & $0.50$ & $0.04$ & \hspace*{.4cm} $  987$ & $   88$ & $73.2$ & $ 7.9$ & $117.9$ & $10.5$ & $-68.5$ & $11.0$ \\
WD\,1448+077 & $0.59$ & $0.04$ & \hspace*{.4cm} $-1121$ & $  144$ & $-151.1$ & $ 5.5$ & $-132.6$ & $17.1$ & $-9.6$ & $ 4.7$ \\
WD\,1457$-$086 & $0.08$ & $0.02$ & \hspace*{.4cm} $ 1793$ & $   75$ & $18.9$ & $ 6.5$ & $213.2$ & $ 8.9$ & $-13.3$ & $ 6.9$ \\
WD\,1515$-$164 & $0.23$ & $0.04$ & \hspace*{.4cm} $ 2101$ & $   86$ & $57.0$ & $ 6.5$ & $249.6$ & $10.2$ & $-8.9$ & $ 8.8$ \\
WD\,1524$-$749 & $0.49$ & $0.02$ & \hspace*{.4cm} $-1388$ & $  119$ & $-136.9$ & $15.4$ & $-167.3$ & $14.5$ & $-16.1$ & $10.4$ \\
WD\,1537$-$152 & $0.12$ & $0.04$ & \hspace*{.4cm} $ 1752$ & $  121$ & $-28.2$ & $ 8.0$ & $208.2$ & $14.4$ & $23.9$ & $12.5$ \\
WD\,1555$-$089 & $0.24$ & $0.02$ & \hspace*{.4cm} $ 1549$ & $   52$ & $55.1$ & $ 4.1$ & $183.4$ & $ 6.1$ & $20.3$ & $ 5.1$ \\
WD\,1609+135 & $0.23$ & $0.01$ & \hspace*{.4cm} $ 1677$ & $   23$ & $65.5$ & $ 3.9$ & $197.6$ & $ 2.7$ & $ 6.2$ & $ 3.4$ \\
WD\,1636+057 & $0.26$ & $0.01$ & \hspace*{.4cm} $ 1395$ & $   31$ & $23.1$ & $ 4.6$ & $164.5$ & $ 3.6$ & $ 9.4$ & $ 3.1$ \\
WD\,1834$-$781 & $0.48$ & $0.03$ & \hspace*{.4cm} $  944$ & $   49$ & $-55.4$ & $ 7.2$ & $111.5$ & $ 5.9$ & $-93.2$ & $ 6.3$ \\
WD\,1952$-$206 & $0.37$ & $0.03$ & \hspace*{.4cm} $ 1161$ & $   51$ & $32.9$ & $ 2.9$ & $137.3$ & $ 6.0$ & $-64.1$ & $ 5.1$ \\
WD\,1959+059 & $0.29$ & $0.02$ & \hspace*{.4cm} $ 1343$ & $   33$ & $29.8$ & $ 4.2$ & $158.8$ & $ 3.9$ & $34.4$ & $ 3.4$ \\
WD\,2007$-$219 & $0.23$ & $0.01$ & \hspace*{.4cm} $ 1484$ & $   22$ & $-35.5$ & $ 2.6$ & $175.0$ & $ 2.6$ & $ 9.4$ & $ 2.3$ \\
WD\,2014$-$575 & $0.12$ & $0.01$ & \hspace*{.4cm} $ 1763$ & $   29$ & $32.4$ & $ 3.1$ & $208.3$ & $ 3.4$ & $-27.7$ & $ 3.3$ \\
WD\,2139+115 & $0.19$ & $0.02$ & \hspace*{.4cm} $ 1769$ & $   27$ & $-54.5$ & $ 4.9$ & $209.5$ & $ 3.1$ & $-62.3$ & $ 4.9$ \\
WD\,2151$-$307 & $0.03$ & $0.02$ & \hspace*{.4cm} $ 1858$ & $   48$ & $ 3.9$ & $ 5.3$ & $219.9$ & $ 5.7$ & $-7.2$ & $ 4.7$ \\
WD\,2254+126 & $0.15$ & $0.02$ & \hspace*{.4cm} $ 1706$ & $   30$ & $-37.6$ & $ 5.4$ & $201.0$ & $ 3.5$ & $-9.7$ & $ 3.8$ \\
WD\,2306+124 & $0.10$ & $0.01$ & \hspace*{.4cm} $ 2057$ & $   26$ & $ 9.2$ & $ 3.3$ & $242.0$ & $ 3.1$ & $ 4.4$ & $ 3.2$ \\
WD\,2311$-$260 & $0.14$ & $0.08$ & \hspace*{.4cm} $ 1758$ & $  248$ & $ 2.4$ & $29.6$ & $210.2$ & $29.7$ & $-26.9$ & $12.8$ \\
WD\,2318$-$226 & $0.20$ & $0.06$ & \hspace*{.4cm} $ 1531$ & $  123$ & $30.4$ & $14.1$ & $181.1$ & $14.6$ & $-15.1$ & $ 6.5$ \\
WD\,2322$-$181 & $0.33$ & $0.04$ & \hspace*{.4cm} $ 1578$ & $   67$ & $-92.4$ & $11.7$ & $186.5$ & $ 7.8$ & $-30.1$ & $ 5.0$ \\
WD\,2326+049 & $0.19$ & $0.01$ & \hspace*{.4cm} $ 1974$ & $   14$ & $55.7$ & $ 2.4$ & $232.1$ & $ 1.6$ & $-6.0$ & $ 2.0$ \\
WD\,2329$-$332 & $0.13$ & $0.05$ & \hspace*{.4cm} $ 2080$ & $  105$ & $-6.8$ & $12.7$ & $246.1$ & $12.5$ & $-11.4$ & $ 5.0$ \\
WD\,2333$-$049 & $0.23$ & $0.03$ & \hspace*{.4cm} $ 1914$ & $   60$ & $70.2$ & $ 8.5$ & $225.1$ & $ 7.0$ & $-0.8$ & $ 4.6$ \\
WD\,2333$-$165 & $0.06$ & $0.01$ & \hspace*{.4cm} $ 1831$ & $   31$ & $17.8$ & $ 3.7$ & $215.5$ & $ 3.7$ & $-45.8$ & $ 2.5$ \\
WD\,2347$-$192 & $0.18$ & $0.04$ & \hspace*{.4cm} $ 2026$ & $  104$ & $47.9$ & $12.2$ & $238.8$ & $12.3$ & $ 2.3$ & $ 4.7$ \\
WD\,2348$-$244 & $0.23$ & $0.02$ & \hspace*{.4cm} $ 1621$ & $   36$ & $62.1$ & $ 4.2$ & $190.9$ & $ 4.3$ & $-26.9$ & $ 3.0$ \\
WD\,2349$-$283 & $0.13$ & $0.03$ & \hspace*{.4cm} $ 1633$ & $   64$ & $ 9.5$ & $ 6.6$ & $192.5$ & $ 7.5$ & $-14.2$ & $ 2.8$ \\
WD\,2350$-$248 & $0.08$ & $0.03$ & \hspace*{.4cm} $ 1974$ & $   64$ & $16.4$ & $ 7.6$ & $232.7$ & $ 7.5$ & $-2.1$ & $ 4.6$ \\
WD\,2351$-$368 & $0.85$ & $0.07$ & \hspace*{.4cm} $  340$ & $  154$ & $77.8$ & $ 8.0$ & $40.1$ & $18.1$ & $60.5$ & $ 3.7$ \\
WD\,2354$-$151 & $0.15$ & $0.07$ & \hspace*{.4cm} $ 1705$ & $  183$ & $28.3$ & $22.2$ & $200.8$ & $21.6$ & $-6.6$ & $ 7.6$ \\
\hline
\end{tabular}
\end{table*}

\subsection{The Monte Carlo error propagation code\label{err}}
An error propagation code based on a Monte Carlo method \citep{gershenfeld99} 
was developed in order to take errors into account. 
Let $x_0$ be an input parameter of a numerical code and $\sigma$ 
its corresponding error. The error of the output parameter $z$ 
is obtained as follows: first a large number of representative 
(e.g.\ 1\,000) $x$-values obeying a Gaussian distribution 
(with $x_0$ as mean value and $\sigma$ as standard deviation) is chosen. 
Then $z$ is computed for each $x$ and the mean value and 
the error of $z$ are calculated. 
Instead of using only one input parameter it is possible to deal with 
several input parameters which are varied simultaneously. 
This method can be useful in cases where a direct error propagation 
cannot be carried out due to the complexity of the problem.

We applied our error propagation code to the \citet{odenkirchen92} code 
described in Sect. \ref{code} for calculating errors of the 
kinematic parameters of the white dwarfs in our sample. 
Input parameters were radial velocities, proper motion 
components and spectroscopic distances together with 
their corresponding errors. 
Then $1\,000$ representative values of $v_{\rm rad}$, 
$\mu_{\alpha}{\rm cos}\delta$, $\mu_{\delta}$ and $d$ obeying
a Gaussian distribution were chosen simultaneously 
and the output parameters together with their errors were computed. 
\section{Our kinematic analysis method\label{method}}
Coordinates $\alpha$ and $\delta$, proper motion components, distances and
radial velocities allow us to benefit from the full 3D information on 
Galactic positions and velocities of the white dwarfs. 
These quantities can be transformed to the 
Galactic system to obtain the coordinates $X$,$Y$, and $Z$ and the 
velocity components $U$(in the Galactic disk 
in direction towards the Galactic centre), $V$(in the direction of the 
Galactic rotation) and $W$(perpendicular to the Galactic disk). 
We adopt the IAU standard \citep{kerr86} of $R_0=8.5\,\rm{kpc}$ 
for the distance of the
sun from the Galactic centre and $V(R_0)=220\,\rm{km\,s^{-1}}$ for the
rotational velocity at the sun's position. For the solar peculiar motion we
take the values of \cite{dehnen98}: $U_{\rm solar}=10\,\rm{km\,s^{-1}}$, 
$V_{\rm solar}=5.25\,\rm{km\,s^{-1}}$ 
and $W_{\rm solar}=7.17\,\rm{km\,s^{-1}}$.

Having calculated $U$, $V$, and $W$ the classical analysis 
in the $U$\/-$V$-velocity diagram 
yields information on population membership. 
This is possible because we have a local sample. 
Furthermore we make use of more sophisticated analysis tools 
(see Sect. \ref{jzecc} and Sect. \ref{orbits}).
Given a Galactic potential the observed values of $X$, $Y$, $Z$ and 
$U$, $V$, $W$ can be used to integrate the equation of motion of the star 
and to follow the dynamical evolution of these quantities over time.
The result is the orbit of the star in the Galaxy. 
At this point it should be noted that this orbit 
cannot be taken as real but is an idealised representation. 
Taken the initial conditions of a white dwarf at the present time it cannot 
necessarily be extrapolated where it will be after a certain time has elapsed,
because the mean-field approximation 
for the Galactic potential is simplified and scattering processes 
between individual stars are neglected.    
But the advantage of this approach is that orbital parameters 
(such as the eccentricity or the angular momentum) can be 
computed which allow us to gain additional 
information on the population-membership of the white dwarfs. 
\subsection{Calculation of orbits and kinematic parameters\label{code}}
For the computation of orbits and kinematic parameters we used a code 
by \citet{odenkirchen92} based on a Galactic potential by \citet{allen91}.
This potential is completely analytical, symmetrical with respect 
to the $Z$-axis, 
time-independent and fitted to the observational values 
of the Galactic rotation curve and the perpendicular force. 
For the integration of the equations of motion \citet{odenkirchen92} 
implemented the Bulirsch-Stoer extrapolation method with adaptive step-size. 
Input parameters are $\alpha(2000)$, 
$\delta(2000)$, $d$, $v_{\rm rad}$, $\mu_{\alpha}{\rm cos}\delta$, 
$\mu_{\delta}$.
The motion of the star was followed for a time span of $2\,{\rm Gyr}$ with a 
time-step of $0.001\,{\rm Gyr}$. 

$X$, $Y$, $Z$ and $U$, $V$, $W$ were 
calculated as a function of time. 
Output parameters are the total energy   
$E$ and $z$-component of angular momentum $J_Z$, 
orbital quantities like the perigalactic distance $R_{\rm min}$,
the apogalactic distance $R_{\rm max}$, the eccentricity 
 $e={{R_{\rm max}-R_{\rm min}}\over {R_{\rm max}+R_{\rm min}}}$ and 
the maximal height above the Galactic disk $z_{\rm max}$. 

In order to check if our results depend on the choice of a special 
Galactic potential 
we recalculated orbits and kinematic parameters with another potential taken 
from \citet{flynn96} and compared them to those obtained with the 
potential of \citet{allen91}. 
We found that the changes are small and do not affect our classification 
criteria for distinction of the different populations. 
While $J_Z$ is not altered at all, the eccentricity typically reduces 
by only $0.01$ and the character of the orbits (thin disk, thick disk or 
halo type) stays the same though the orbits are changed slightly. 
This provides a justification that our results do not depend 
sensitively on the choice of a particular potential.    
\subsection{The calibration sample\label{cal}}
Unlike for main-sequence stars the population membership of white dwarfs cannot be determined 
from spectroscopically measured metallicities. 
Therefore we have to rely on kinematic criteria. 
Those criteria have to be calibrated using a suitable calibration sample of 
main-sequence stars. 
In our case this sample consists of F and G main-sequence stars from 
\citet{edvard93}, \citet{fuhrmann98}, and 
Fuhrmann (2000, {\tt http://www.xray.mpe.mpg.de/fuhrmann/}). 

For both samples a detailed abundance analysis has been carried out.  
Fuhrmann found that the disk and halo populations can be distinguished best 
in the [Mg/Fe] versus [Fe/H] diagram. Halo and thick disk stars can be 
separated by means of their [Fe/H] abundances. Halo and thick disk stars 
possess a higher [Mg/Fe] ratio than thin disk stars 
($\alpha$ Process enhanced).    

\begin{figure}
  \resizebox{\hsize}{!}{
\begin{psfrags}
\psfrag{x}{${\rm Fe}/{\rm H}$}
\psfrag{y}{${\rm Mg}/{\rm Fe}$}
\psfrag{thin disk}{thin disk}
\psfrag{thick disk}{thick disk}
\psfrag{halo}{halo}
    \includegraphics[width=17cm]{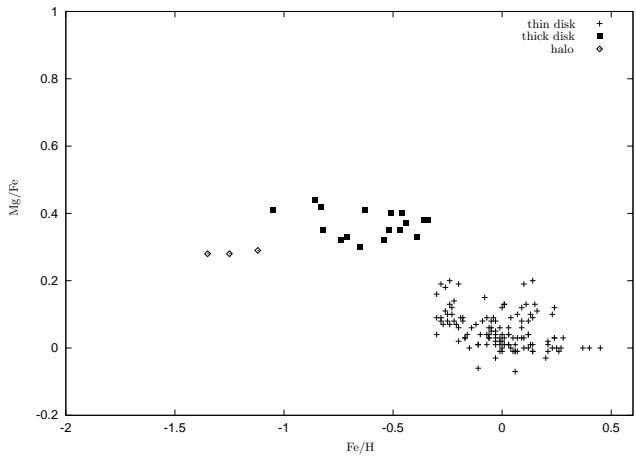}
\end{psfrags}
}
  \caption{[Mg/Fe]-[Fe/H] abundance-diagram for the main-sequence stars of the calibration sample}
\label{met}
\end{figure}

We selected a subsample of $137$ from the Fuhrmann and Edvardsson stars 
for which distances and proper motion are available at 
the ARI Database for Nearby Stars 
({\tt http://www.ari.uni-heidelberg.de/aricns}).
Radial velocities were obtained  from
the ARI Database as well  
or the compilation of \citet{barbier94}. 
Those parameters are necessary to compare 
the kinematics of the main-sequence stars with those of the white dwarfs. 
In Fig.~\ref{met} the [Mg/Fe] versus [Fe/H] abundances for the $137$ 
main-sequence stars are shown. 
Those stars are divided into halo, thick disk and
thin disk according to their position in the diagram following Fuhrmann 
(2000). 
The halo stars have $[{\rm Fe}/{\rm H}]<-1.2$, 
the thick disk stars  $-1.1\le[{\rm Fe}/{\rm H}]\le-0.3$ and 
$[{\rm Mg}/{\rm Fe}]\ge 0.3$ and the thin disk stars 
$[{\rm Fe}/{\rm H}]>-0.3$ and $[{\rm Mg}/{\rm  Fe}] \le 0.2$. 
Stars in the overlapping  area between the thin and the thick disk were 
neglected in order to ensure a clear 
distinction between the two disk populations. 
Due to the low number of halo and thick disk stars in the calibration sample 
the separation between these two population is somewhat arbritrary, 
but as will be demonstrated later, halo and thick disk stars 
show very distinct kinematic 
properties so that they cannot be confused easily. 

The star HD\,148816 is classified as thick disk star 
by means of its abundances ($[{\rm Fe/H}]=-0.74$, $[{\rm Mg/Fe}]=0.32$).  
However it will be demonstrated later in Sect.~\ref{uv} and 
Sect.~\ref{jzecc} that its
kinematics are not compatible with a thick disk but a halo member.   
Therefore we will exclude this star from the calibration sample.  
\section{Kinematic studies \label{stud}}
We calculated orbits and kinematic parameters for all white dwarfs 
(see Table~\ref{kinpar}) and
all main-sequence stars for intercomparison. 
For our white dwarf sample the errors of $e$, $J_Z$, $U$, $V$, $W$ were 
computed with our Monte Carlo error propagation code 
described in Sect.~\ref{err}. They can be found in Table~\ref{kinpar} as well. 
\subsection{The $U$\/-$V$-velocity diagram\label{uv}}
\begin{figure*}
  \centering
\begin{psfrags}
\psfrag{x}{$V/{\rm km\,s^{-1}}$}
\psfrag{y}{$U/{\rm km\,s^{-1}}$}
\psfrag{thin disk}{thin disk}
\psfrag{thick disk}{thick disk}
\psfrag{halo}{halo}
\psfrag{1sigma}{$1\sigma$-limit}
\psfrag{2sigma}{$2\sigma$-limit}
\psfrag{HD148816}{HD\,148816}
\psfrag{HD284248}{HD\,284248}
\psfrag{HD194598}{HD\,194598}
\psfrag{HD221830}{HD\,221830}
\psfrag{HD165401}{HD\,165401}
\psfrag{HD400}{HD\,400}
    \includegraphics[width=17cm]{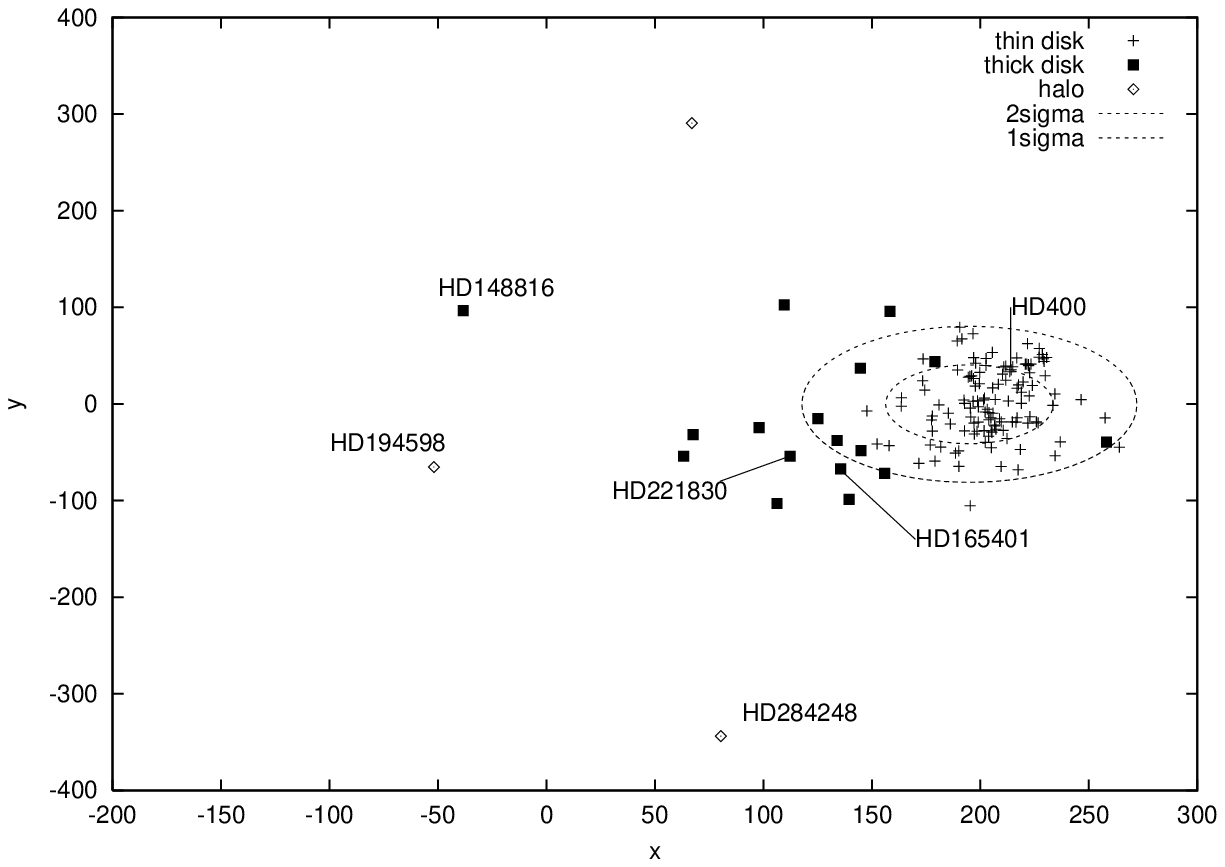}
\end{psfrags}
  \caption{$U$\/-$V$-velocity diagram for the main-sequence stars, 
dashed lines: $1\sigma$-, $2\sigma$-contours, symbols with numbers are stars mentioned in the text}
\label{uvms}
  \centering
\begin{psfrags}
\psfrag{x}{$V/{\rm km\,s^{-1}}$}
\psfrag{y}{$U/{\rm km\,s^{-1}}$}
\psfrag{white dwarfs}{white dwarfs}
\psfrag{thick disk}{thick disk}
\psfrag{halo}{halo}
\psfrag{1sigma}{$1\sigma$-limit}
\psfrag{2sigma}{$2\sigma$-limit}
\psfrag{wd1426-276}{WD\,1426$-$276}
\psfrag{wd0013-241}{WD\,0013$-$241}
\psfrag{he1152-1244}{HE\,1152$-$1244}
\psfrag{wd1448+077}{WD\,1448+077}
\psfrag{wd1524-749}{WD\,1524$-$749}
\psfrag{wd0252-350}{WD\,0252$-$350}
\psfrag{wd2351-368}{WD\,2351$-$368}
    \includegraphics[width=17cm]{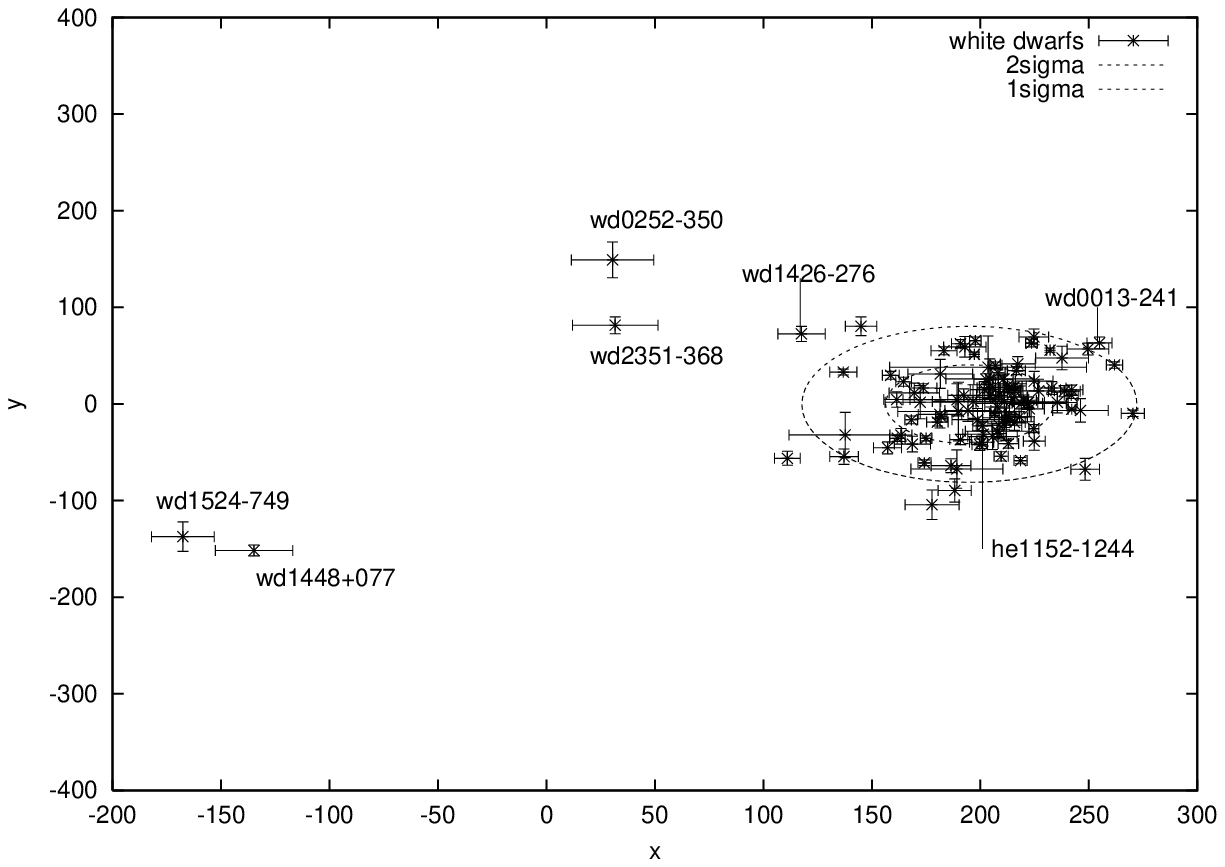}
\end{psfrags}
  \caption{$U$\/-$V$-velocity diagram for the white dwarfs with 
$1\sigma$-, $2\sigma$-contours from Fig.~\ref{uvms}, symbols with numbers are white dwarfs mentioned in the text}
  \label{uvwd}
\end{figure*}

A classical tool for kinematic investigations is the
$U$\/-$V$-velocity diagram. In Fig.~\ref{uvms} $U$ is plotted  
versus $V$ for the main-sequence stars. 
Thin disk stars are marked by crosses, thick disk stars by squares, 
and halo stars by diamonds. 
For all disk stars the mean values and standard
deviations of the two velocity components have been calculated: 
$\left<U_{\rm ms}\right>=-0.3\,\rm{km\,s^{-1}}$, $\left<V_{\rm ms}\right>=195\,\rm{km\,s^{-1}}$, 
$\sigma_{U_{\rm ms}}=41\,\rm{km\,s^{-1}}$, $\sigma_{V{\rm ms}}=39\,\rm{km\,s^{-1}}$.
The $1\sigma$ and $2\sigma$-contours are shown in Fig.~\ref{uvms}. 
\cite{chiba00} found $\sigma_{V}=50\,\rm{km\,s^{-1}}$,
$\sigma_{U}=46\,\rm{km\,s^{-1}}$ for the thick disk.  
Most of the thin disk stars stay within the $2\sigma$-contours. 
The area outside the $2\sigma$-limit contains all halo 
and many of the thick disk stars. 
The three halo main-sequence stars are characterised by values of 
$\sqrt{U^2+(V-195)^2}\ge 150\,\rm{km\,s^{-1}}$. 
As already mentioned we ignore the thick disk star HD\,148816 
for kinematic studies. It stands out by its peculiar position 
in the $U$\/-$V$-diagram with a negative value of 
$V$. 

In Fig.~\ref{uvwd} the $U$\/-$V$-plot is shown for the white dwarfs with error
bars. 
For comparison the $1\sigma$ and $2\sigma$-contours derived 
for the main-sequence stars appear as well. 
In order to detect halo and thick disk stars we concentrate on the
14 white dwarfs outside the $2\sigma$-limit. 
Four stars have $\sqrt{U^2+(V-195)^2}\ge 150\,\rm{km\,s^{-1}}$. 
WD\,0252$-$350 and WD\,2351$-$368 have a $V$ velocity around $30\,\rm
{km\,s^{-1}}$ which is clearly below the $220\,\rm{km\,s^{-1}}$ of the
local standard of rest and their $U$ values are high. This indicates
that they do not rotate with the Galactic disk and can move far away
from the Galactic centre. 
This is characteristic for halo stars. 
WD\,1448+077 and WD\,1524$-$749 do even have a negative value of the 
$V$-velocity, 
which means that their orbits are retrograde. 
They therefore do not belong to the disk. 
We also found a main-sequence star (HD\,194598) on a retrograde orbit which
 according to its abundance pattern is a member of the halo.   
The other $10$ white dwarfs outside the $2\sigma$-contours are possible 
thick disk candidates belonging to the high velocity tail (large deviations 
in $U$ and $V$ from the disk mean values) of the disk. 
\subsection{The $J_Z$-eccentricity-diagram\label{jzecc}}
\begin{figure*}
  \centering
\begin{psfrags}
\psfrag{x}{$e$}
\psfrag{y}{$J_Z/{\rm kpc\,km\,s^{-1}}$}
\psfrag{thin disk}{thin disk}
\psfrag{thick disk}{thick disk}
\psfrag{halo}{halo}
\psfrag{Region1}{Region~1}
\psfrag{Region2}{Region~2}
\psfrag{Region3}{Region~3}
\psfrag{Region4}{Region~4}
\psfrag{HD148816}{HD\,148816}
\psfrag{HD284248}{HD\,284248}
\psfrag{HD194598}{HD\,194598}
\psfrag{HD221830}{HD\,221830}
\psfrag{HD165401}{HD\,165401}
\psfrag{HD400}{HD\,400}
    \includegraphics[width=17cm]{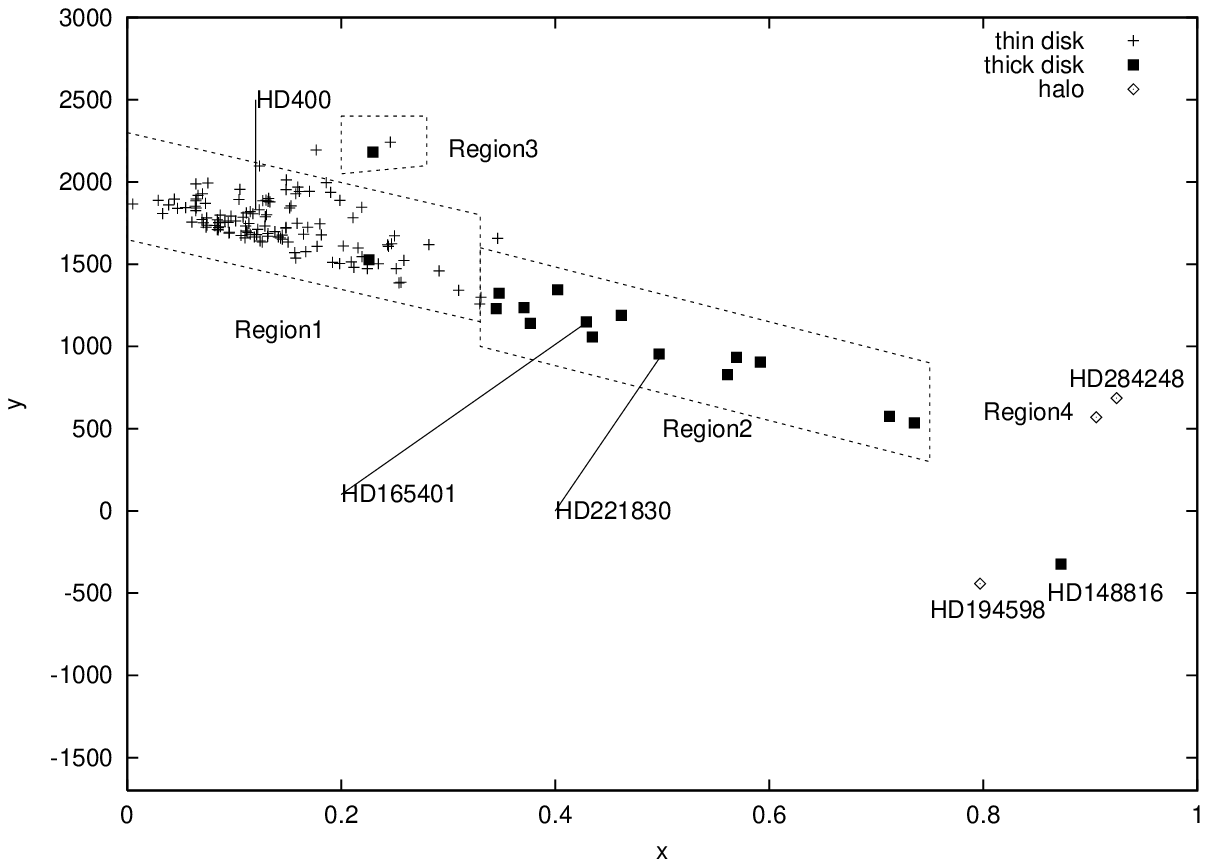}
\end{psfrags}
  \caption{$J_Z$-$e$-diagram for the main-sequence stars, symbols with numbers are stars mentioned in the text}
  \label{ecc1}
  \centering
\begin{psfrags}
\psfrag{x}{$e$}
\psfrag{y}{$J_Z/{\rm kpc\,km\,s^{-1}}$}
\psfrag{white dwarfs}{white dwarfs}
\psfrag{thick disk}{thick disk}
\psfrag{halo}{halo}
\psfrag{Region1}{Region~1}
\psfrag{Region2}{Region~2}
\psfrag{Region3}{Region~3}
\psfrag{Region4}{Region~4}
\psfrag{wd1426-276}{WD\,1426$-$276}
\psfrag{wd0013-241}{WD\,0013$-$241}
\psfrag{he1152-1244}{HE\,1152$-$1244}
\psfrag{wd1448+077}{WD\,1448+077}
\psfrag{wd1524-749}{WD\,1524$-$749}
\psfrag{wd0252-350}{WD\,0252$-$350}
\psfrag{wd2351-368}{WD\,2351$-$368}
    \includegraphics[width=17cm]{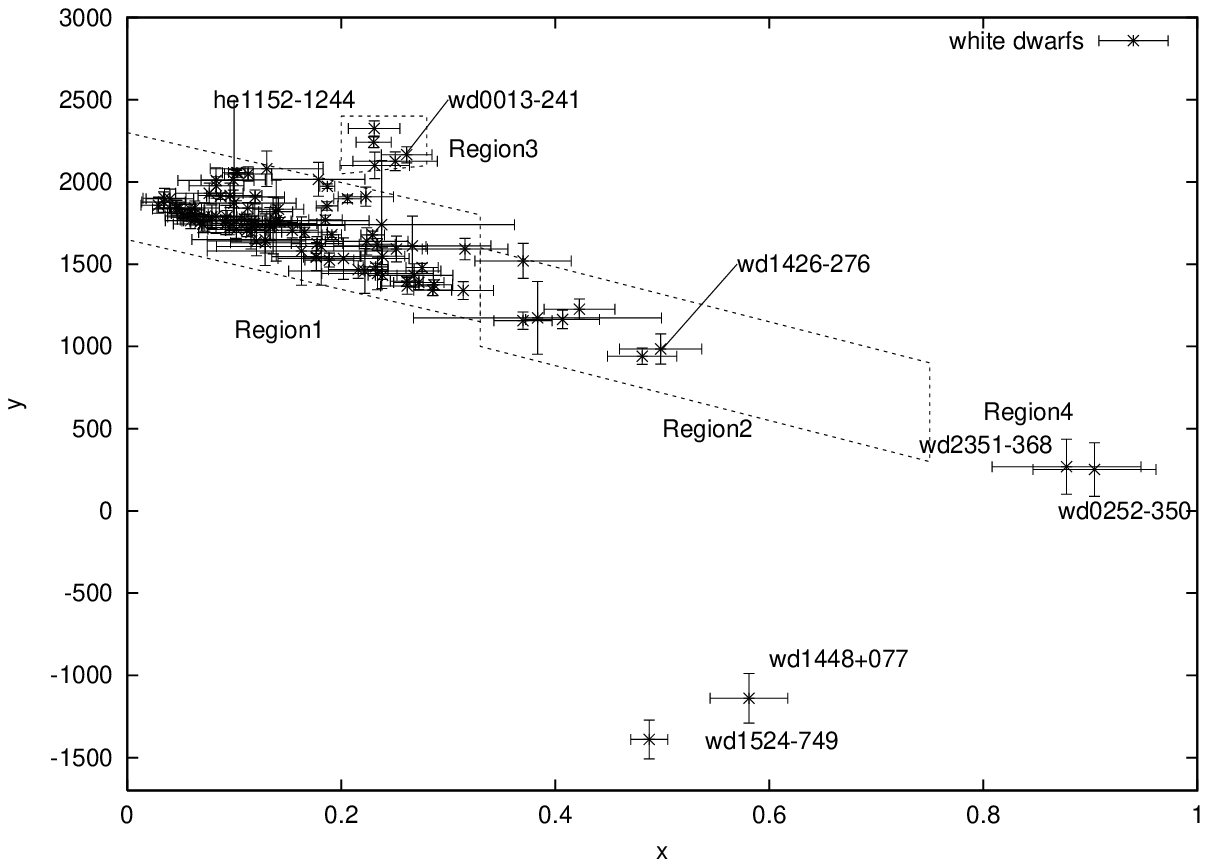}
\end{psfrags}
  \caption{$Jz$-$e$-diagram of the white dwarfs, symbols with numbers are white dwarfs mentioned in the text}
  \label{ecc2}
\end{figure*}
The $U$\/-$V$-plot is not the only source of information about population 
membership. 
Two important orbital parameters
are the $z$-component of the angular momentum $J_Z$ and the
eccentricity of the orbit $e$. Both are plotted against each other
for the main-sequence stars in Fig.~\ref{ecc1}. 
The different populations can be distinguished well in this diagram. 
The thin disk stars cluster in an area of
low eccentricity and $J_Z$ around $1\,800\,\rm{kpc \, km\,s^{-1}}$,
which we denote by Region~1. 

The thick disk stars possess higher
eccentricities and lower angular momenta. They can be found in
Region~2. 
Two stars, one from the thin disk and one from the thick disk, 
populate an intermediate Region~3 with 
eccentricity around $e=0.22$ and large $J_Z$. 
The reason for defining this additional Region~3 is that (as will be seen in Fig.~\ref{ecc2}) 
five white dwarfs that do neither belong to Region~1 nor to Region~2 
can be found there.  
The halo stars with very high eccentricity and smaller
$J_Z$ can be found in Region~4. 
This region also contains the star HD\,148816 which 
suggests again that this star really belongs to the halo and not 
to the thick disk. 

In Fig.~\ref{ecc2} the $J_Z$-$e$-diagram for the white dwarfs is shown 
with the different regions as defined above. 
Let us first concentrate on the four halo white dwarf candidates. 
WD\,0252$-$350 and WD\,2351$-$368 are the stars with the largest 
eccentricity, they lie very close to the halo main-sequence stars. 
WD\,1448+077 and WD\,1524$-$749 again populate an exceptional region in the
diagram. 
Their value of $J_Z$ is negative. 
Their eccentricity is not as large as that of the other halo stars. 

Five of the ten thick disk candidates lying outside the $2\sigma$-limit 
in the $U$\/-$V$-velocity diagram are present in Region~2. 
Two stars, WD\,1952$-$206 and WD\,0204$-$233 did not appear outside 
the $2\sigma$-limit, but they are clearly situated in Region~2.

Five white dwarfs can be found in Region~3. Four of them 
have already been classified as possible thick disk members 
in the $U$\/-$V$-plot, 
only WD\,0509$-$007 is a new thick disk candidate.

Of the 10 thick disk candidates in the $U$\/-$V$-plot, 9 are present in 
Regions 2 and 3. Three additional thick disk candidates only show up 
in the $J_Z$-$e$-diagram. 
Hence the $U$\/-$V$-velocity plot alone is not sufficient to 
decide on population membership. 
The $J_Z$-$e$-diagram can provide further information.
\subsection{Orbits \label{orbits}}
The eccentricity was extracted from the orbit of the white dwarfs. 
It can also be interesting to look at the orbits themselves. 
We consider the motion perpendicular to the plane (in the $Z$-direction) 
as a function of the motion inside the plane in radial direction 
($\rho=\sqrt{X^2+Y^2}$).
This is called the meridional plot.
We compare meridional plots of main-sequence and white dwarf stars of the 
different populations. 
For ease of comparison the scale is the same for all plots. 

\begin{figure}
  \resizebox{\hsize}{!}{
    \begin{psfrags}
      \psfrag{rho/kpc}{$\rho/{\rm kpc}$}
      \psfrag{Z/kpc}{$Z/{\rm kpc}$}
      \includegraphics{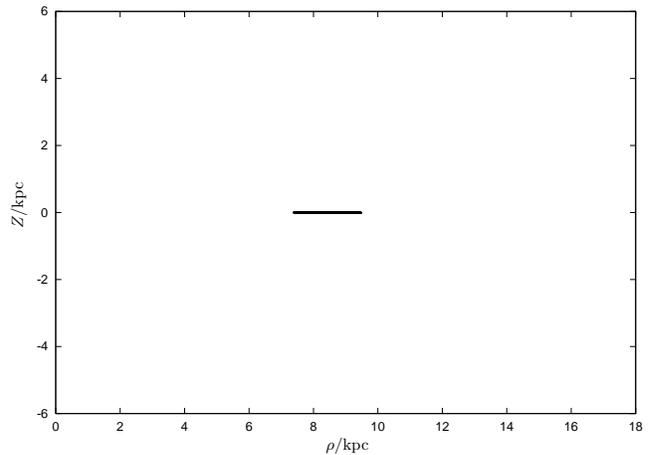}
    \end{psfrags}
    }
  \caption{Orbit of HD\,400, a thin disk main-sequence star}
  \label{HD400}
\end{figure}

\begin{figure}
  \resizebox{\hsize}{!}{
    \begin{psfrags}
      \psfrag{rho/kpc}{$\rho/{\rm kpc}$}
      \psfrag{Z/kpc}{$Z/{\rm kpc}$}
      \includegraphics{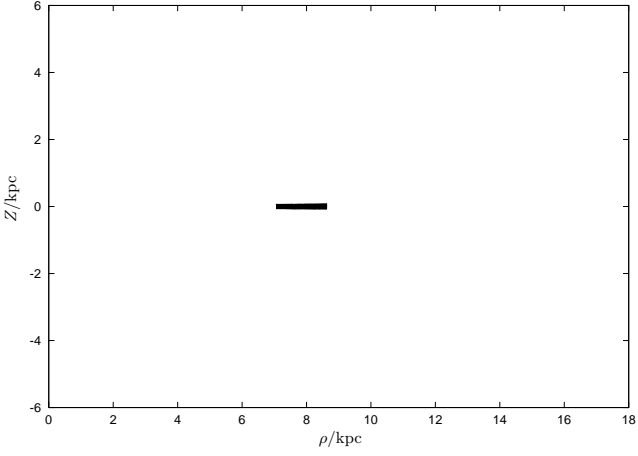}
    \end{psfrags}
    }
  \caption{Orbit of HE\,1152$-$1244, a thin disk white dwarf}
  \label{he1152}
\end{figure}

We start with the thin disk stars. 
Fig.~\ref{HD400} shows the orbit of HD\,400, a typical 
main-sequence thin disk star. 
It is situated in Region~1 in the $J_Z$-$e$-diagram and its 
extensions in the $\rho$ and the $Z$-directions are small. 
The white dwarf HE\,1152$-$1244 (Fig.~\ref{he1152}) has an orbit 
similar to that of HD\,400 suggesting that it belongs to the thin disk. 
Actually most of our white dwarfs possess thin disk meridional plots 
similar to this one.

\begin{figure}
  \resizebox{\hsize}{!}{
    \begin{psfrags}
      \psfrag{rho/kpc}{$\rho/{\rm kpc}$}
      \psfrag{Z/kpc}{$Z/{\rm kpc}$}
      \includegraphics{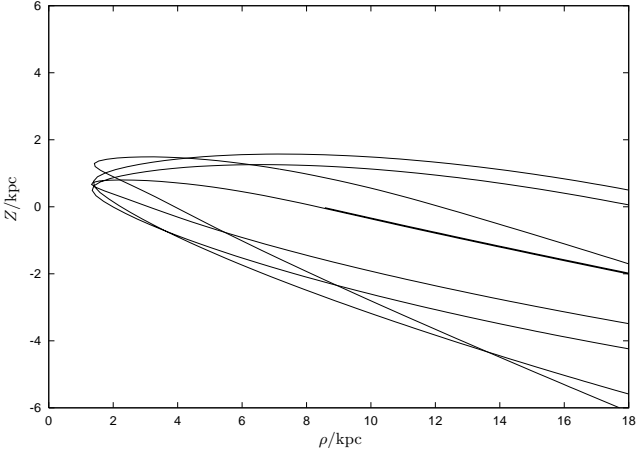}
    \end{psfrags}
    }
  \caption{Orbit of HD\,284248, a halo main-sequence star}
  \label{HD284248}
\end{figure}

\begin{figure}
  \resizebox{\hsize}{!}{
    \begin{psfrags}
      \psfrag{rho/kpc}{$\rho/{\rm kpc}$}
      \psfrag{Z/kpc}{$Z/{\rm kpc}$}
      \includegraphics{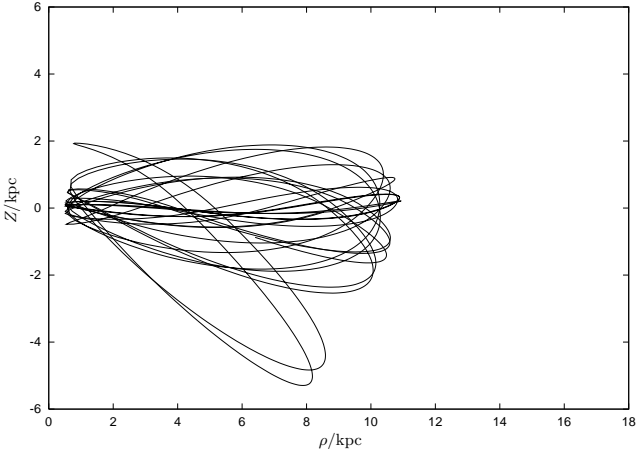}
    \end{psfrags}
    }
  \caption{Orbit of WD\,0252$-$350, a halo white dwarf}
  \label{wd0252}
\end{figure}

\begin{figure}
  \resizebox{\hsize}{!}{
    \begin{psfrags}
      \psfrag{rho/kpc}{$\rho/{\rm kpc}$}
      \psfrag{Z/kpc}{$Z/{\rm kpc}$}
      \includegraphics{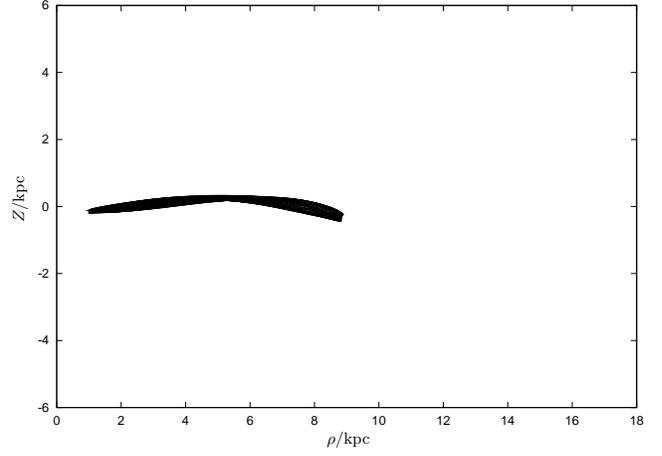}
    \end{psfrags}
    }
  \caption{Orbit of HD\,194598, a retrograde halo main-sequence star}
  \label{HD194598}
\end{figure}

\begin{figure}
  \resizebox{\hsize}{!}{
    \begin{psfrags}
      \psfrag{rho/kpc}{$\rho/{\rm kpc}$}
      \psfrag{Z/kpc}{$Z/{\rm kpc}$}
      \includegraphics{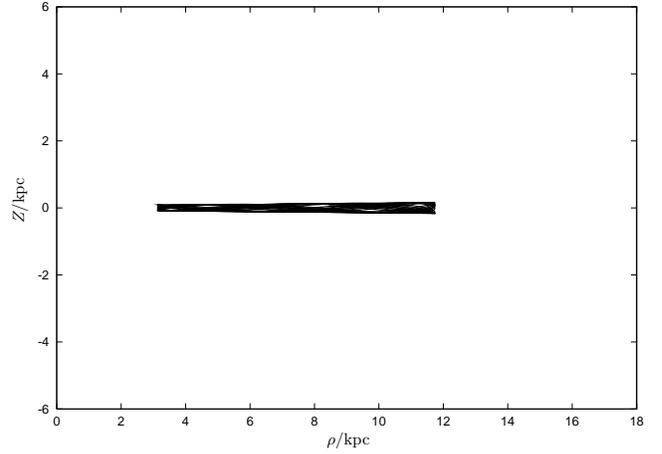}
   \end{psfrags}
    }
  \caption{Orbit of WD\,1448+077, a retrograde halo white dwarf}
  \label{wd1448}
\end{figure}

The orbit of the main-sequence star HD\,284248 (Fig.~\ref{HD284248}) is 
characteristic for a halo object. 
Its extension in $\rho$ is 
so large that it exceeds the range of the plot ($18\,{\rm kpc}$) 
and its vertical distance 
from the Galactic plane is larger than $6\,{\rm kpc}$.  
Due to their orbits the white dwarfs WD\,0252$-$350 (Fig.~\ref{wd0252}) 
and WD\,2351$-$368 also qualify as halo stars. 
The white dwarfs WD\,1448+0077 and WD\,1524$-$749 with 
negative $V$ velocity (retrograde orbit) are compared with the  
main-sequence halo star (HD\,194598) which has a similar $V$-velocity.  
The meridional plot of HD\,194598 
(Fig.~\ref{HD194598}) has smaller vertical and meridional distances 
but a large eccentricity. 
We know from its abundance pattern and 
the $U$\/-$V$-plot and $J_Z$-$e$ plot that it must belong to the halo. 
WD\,1448+0077's (Fig.~\ref{wd1448}) and WD\,1524-749's orbits 
are nearly identical 
to that of HD\,194598. Hence we conclude that they are members of the halo 
population, too. 

\begin{figure}
  \resizebox{\hsize}{!}{
    \begin{psfrags}
      \psfrag{rho/kpc}{$\rho/{\rm kpc}$}
      \psfrag{Z/kpc}{$Z/{\rm kpc}$}
      \includegraphics{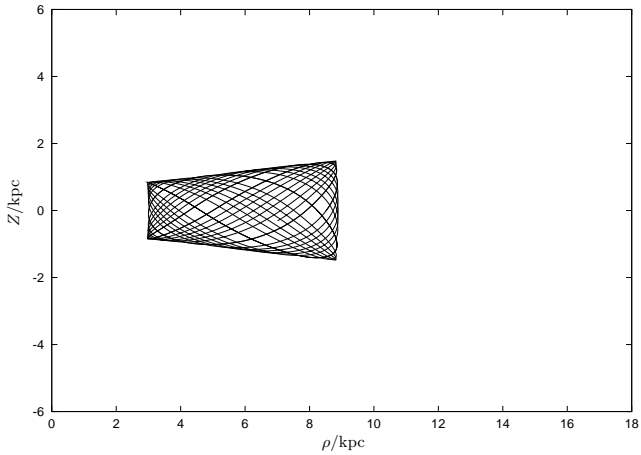}
    \end{psfrags}
    }
  \caption{Orbit of HD\,221830, a thick disk main-sequence star}
  \label{HD221830}
\end{figure}

\begin{figure}
  \resizebox{\hsize}{!}{
    \begin{psfrags}
      \psfrag{rho/kpc}{$\rho/{\rm kpc}$}
      \psfrag{Z/kpc}{$Z/{\rm kpc}$}
      \includegraphics{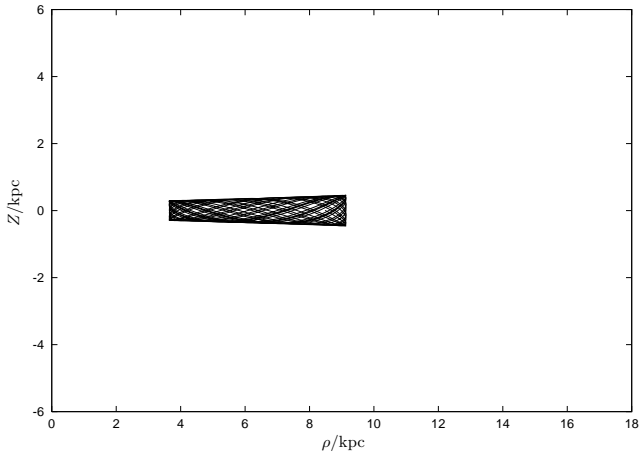}
    \end{psfrags}
    }
  \caption{Orbit of HD\,165401, a thick disk main-sequence star}
  \label{HD165401}
\end{figure}

The thick disk main-sequence stars are characterised by orbits 
more extended in the $\rho$ and $Z$ directions than that of a thin disk star.  
However, they do not cover such a large region in the meridional plot 
as a halo member. 
Examples are the orbits of the thick disk members 
HD\,221830 (Fig.~\ref{HD221830}) and HD\,165401 (Fig.~\ref{HD165401}).   

\begin{figure}
  \resizebox{\hsize}{!}{
    \begin{psfrags}
      \psfrag{rho/kpc}{$\rho/{\rm kpc}$}
      \psfrag{Z/kpc}{$Z/{\rm kpc}$}
      \includegraphics{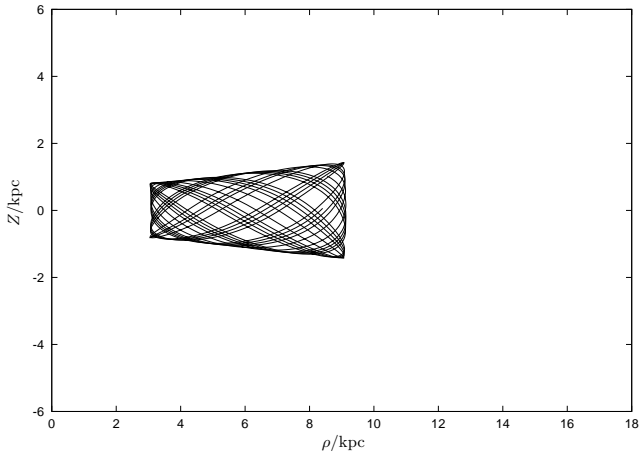}
    \end{psfrags}
    }
  \caption{Orbit of WD\,1426$-$276, a thick disk white dwarf}
  \label{wd1426}
\end{figure}

\begin{figure}
 \resizebox{\hsize}{!}{
   \begin{psfrags}
      \psfrag{rho/kpc}{$\rho/{\rm kpc}$}
      \psfrag{Z/kpc}{$Z/{\rm kpc}$}
      \includegraphics{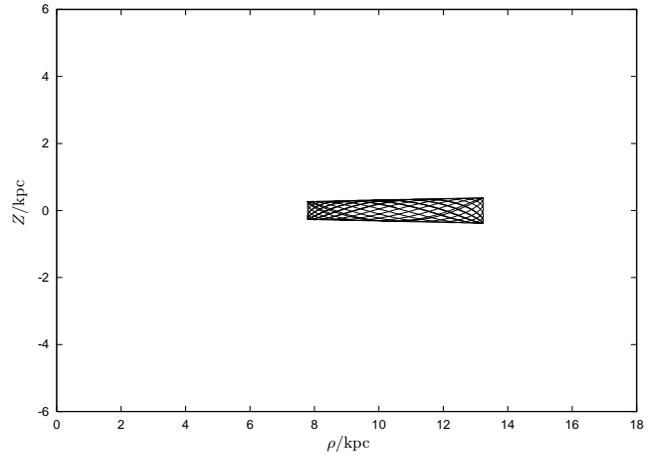}
    \end{psfrags}
    }
  \caption{Orbit of WD\,0013$-$241, a thick disk white dwarf}
  \label{wd0013}
\end{figure}

The next step is to compare those orbits to the orbits of our white dwarf thick disk 
candidates. 
Those are all 13 white dwarfs lying either outside the $2\sigma$-limit 
in the $U$\/-$V$-diagram or in Regions 2 or 3 in the $J_Z$-$e$-diagram.
Four of the thick disk candidates possess meridional plots 
similar to that of HD\,221830.  
As an example the orbit of WD\,1426$-$276 is shown in Fig.~\ref{wd1426}. 
Nine of our candidates possess orbits like HD\,165401. 
A representative is the white dwarf WD\,0013$-$241 (Fig.~\ref{wd0013}).
\begin{table*}
\caption[]
{Classification values for the thick disk candidates\label{classtab}}
\begin{tabular}{crrrrl}
\hline
star & $c_{\rm UV}$ & $c_{\rm J_Z-e}$ & $c_{\rm orb}$ & $c$ & classification\\ 
\hline
HE\,0409$-$5154 & $+1$ & $+1$ & $+1$ & $+3$ & bona fide thick disk\\
HE\,1124+0144 & $+1$ & $+1$ & $+1$ & $+3$ & bona fide thick disk\\
WD\,0013$-$241 & $+1$ & $0$ & $+1$ & $+2$ & bona fide thick disk\\
WD\,0204$-$233 & $-1$ & $+1$ & $+1$ & $+1$ & probable thick disk\\
WD\,0216+143 & $+1$ & $0$ & $+1$ & $+2$ & bona fide thick disk\\
WD\,0509$-$007 & $-1$ & $0$ & $+1$ & $0$ & thin disk\\
WD\,0951$-$155 & $+1$ & $0$ & $+1$ & $+2$ & bona fide thick disk\\
WD\,1334$-$678 & $+1$ & $+1$ & $+1$ & $+3$ & bona fide thick disk\\
WD\,1426$-$276 & $+1$ & $+1$ & $+1$ & $+3$ & bona thick disk\\
WD\,1515$-$164 & $+1$ & $0$ & $+1$ & $+2$ & bona fide thick disk\\
WD\,1834$-$781 & $+1$ & $+1$ & $+1$ & $+3$ & bona fide thick disk\\
WD\,1952$-$206 & $-1$ & $+1$ & $+1$ & $+1$ & probable thick disk\\
WD\,2322$-$181 & $+1$ & $-1$ & $+1$ & $+1$ & probable thick disk\\
\hline
\end{tabular}
\end{table*}
\section{Results\label{results}}
For a refined classification of population membership of our white dwarf 
sample we combine all three criteria discussed above.

Characteristic for halo white dwarfs is a value of  
$\sqrt{U^2+(V-195)^2}\ge 150\,\rm{km\,s^{-1}}$ and 
their location in Region~4 in the $J_Z$-$e$-diagram. 
Their orbits typically have high extensions in $\rho$ and $Z$ like HD\,284248. 
WD\,0252$-$350 and WD\,2351$-$368 fulfil these criteria and are 
therefore classified as halo white dwarfs. 
Stars with retrograde orbits like the halo main-sequence star HD\,194598 
(Fig.~\ref{HD194598}) 
are also members of the halo population. 
The white dwarfs WD\,1448+077 and WD\,1524$-$749 belong to 
this population.    
   
To detect thick disk white dwarfs first all stars either 
situated outside the $2\sigma$-limit 
in the $U$\/-$V$-diagram or in Region~2 or 3 in the $J_Z$-$e$-diagram 
are selected as thick disk candidates. 
According to these criteria 13 white dwarfs of our sample qualify. 

In a second step each candidate is assigned a classification value 
$c$. $c$ is defined as the sum of the individual values $c_{\rm UV}$, 
$c_{\rm J_{Z}e}$ and $c_{\rm orb}$ corresponding to the three 
different criteria: position in $U$\/-$V$-diagram, 
position in $J_Z$-$e$-diagram and orbit. 

We assign $c_{\rm UV}=+1$ to a star outside the $2\sigma$-limit 
in the $U$\/-$V$-diagram, whereas one inside the $2\sigma$-limit 
gets $c_{\rm UV}=-1$. 
The different regions in the $J_Z$-$e$-diagram are characterised by 
$c_{\rm J_{Z}e}=-1$ for Region~1, $0$ for Region~3 and $+1$ for Region~2. 
The third classification value $c_{\rm orb}$ describes the orbits: 
$c=-1$ for orbits of thin disk type and $c=+1$ for orbits of 
thick disk type. 

Then the sum $c=c_{\rm UV}+c_{\rm J_{Z}e}+c_{\rm orb}$ is computed.  
Stars with $c=+3$  or $c=+2$ are considered as bona fide 
thick disk members, those with $c=+1$ as probable thick disk members. 
If $c \le 0$, the star is classified as belonging to the thin disk. 
As we have selected halo members beforehand this classification scheme 
is not applied to halo stars. 

In Table~\ref{classtab} the individual and
combined classification values for the 
thick disk candidates are listed. 
Applying our classification criteria we end up with 
nine bona fide and three probable thick disk white dwarfs. 
The total number of white dwarfs belonging to the thick disk population 
is twelve. 
\subsection{Consistency check for classification criteria \label{consist}}
In this section we perform a consistency check of our classification 
scheme. This is done by applying our kinematic classification criteria to 
our calibration main-sequence sample. 

17 main-sequence stars are known to belong to the thick disk because of 
their abundance levels. 
For reasons mentioned above we exclude the star HD\,184499 
where kinematic and chemical criteria suggest different population 
memberships. 
Hence our total number of thick disk main-sequence stars is 16. 
Ten of them have a classification value $c$ of $+3$ or $+2$ and 
are classified as bona fide thick disk stars. 
Five of them with $c=+1$ are probable thick disk members. 
Only one of them has $c=0$ and is thus misclassified as thin disk star. 
Thus 15 out of 16 thick disk stars are identified correctly. 
This corresponds to a detection efficiency for thick disk 
members of about $94\%$.  
In addition to those 15 stars two thin disk main-sequence star 
also qualify as thick disk 
candidates because they lie outside the $2\sigma$-limit. 
One is classified as probable ($c=+1$), the other as bona fide ($c=+2$) 
thick disk member. 
Therefore both are misidentified as thick disk stars.  
The total number of stars classified as thick disk is 17 including 
two stars which really belong to the thin disk. 
So the contamination with thin disk stars is only about $12\%$. 

We can therefore conclude that our selection criteria are very 
efficient in detecting thick disk stars while the contamination with thin 
disk stars is at an acceptable level.
\subsection{The effects of setting $v_{\rm rad}=0$
\label{vrad0}}
All but two previous kinematic investigations on white dwarfs 
\citep{silvestri01,silvestri02} assumed $v_{\rm rad}=0$ (or $W=0$) 
simply because no radial velocity information was available. 
We test the effects of this assumption by setting 
$v_{\rm rad}$ arbitrarily to zero for all stars in our sample. 

\begin{figure*}
  \centering
  \begin{psfrags}
    \psfrag{x}{$V/{\rm km\,s^{-1}}$}
    \psfrag{y}{$U/{\rm km\,s^{-1}}$}
    \psfrag{vradn}{$v_{\rm rad} \ne 0$}
    \psfrag{vrad0}{$v_{\rm rad}=0$}
    \psfrag{halo}{halo}
    \psfrag{1sigma}{$1\sigma$-limit}
    \psfrag{2sigma}{$2\sigma$-limit}
\psfrag{wd1426-276}{WD\,1426$-$276}
\psfrag{wd0013-241}{WD\,0013$-$241}
\psfrag{he1152-1244}{HE\,1152$-$1244}
\psfrag{wd1448+077}{WD\,1448+077}
\psfrag{wd1524-749}{WD\,1524$-$749}
\psfrag{wd0252-350}{WD\,0252$-$350}
\psfrag{wd2351-368}{WD\.2351$-$368}
    \includegraphics[width=17cm]{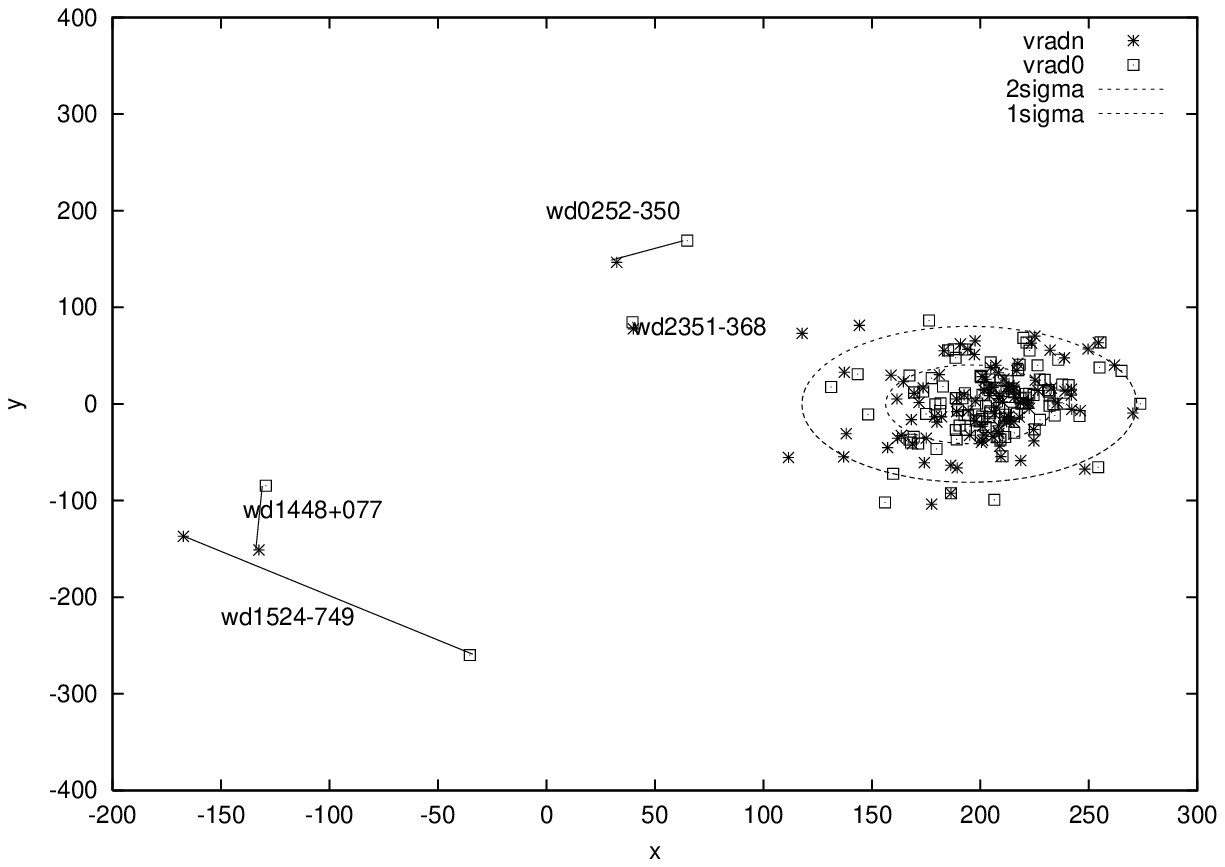}
  \end{psfrags}
  \caption{Effect of setting $v_{\rm rad}=0$ in the $U$\/-$V$-diagram, the change in position of the halo white dwarfs is indicated by lines}
  \label{uvwd0}
  \centering
  \begin{psfrags}
    \psfrag{x}{$e$}
    \psfrag{y}{$J_Z/{\rm kpc\,km\,s^{-1}}$}
    \psfrag{vradn}{$v_{\rm rad} \ne 0$}
    \psfrag{vrad0}{$v_{\rm rad}=0$}
    \psfrag{halo}{halo}
    \psfrag{Region1}{Region~1}
    \psfrag{Region2}{Region~2}
    \psfrag{Region3}{Region~3}
    \psfrag{Region4}{Region~4}
\psfrag{wd1426-276}{WD\,1426$-$276}
\psfrag{wd0013-241}{WD\,0013$-$241}
\psfrag{he1152-1244}{HE\,1152$-$1244}
\psfrag{wd1448+077}{WD\,1448+077}
\psfrag{wd1524-749}{WD\,1524$-$749}
\psfrag{wd0252-350}{WD\,0252$-$350}
\psfrag{wd2351-368}{WD\,2351$-$368}
    \includegraphics[width=17cm]{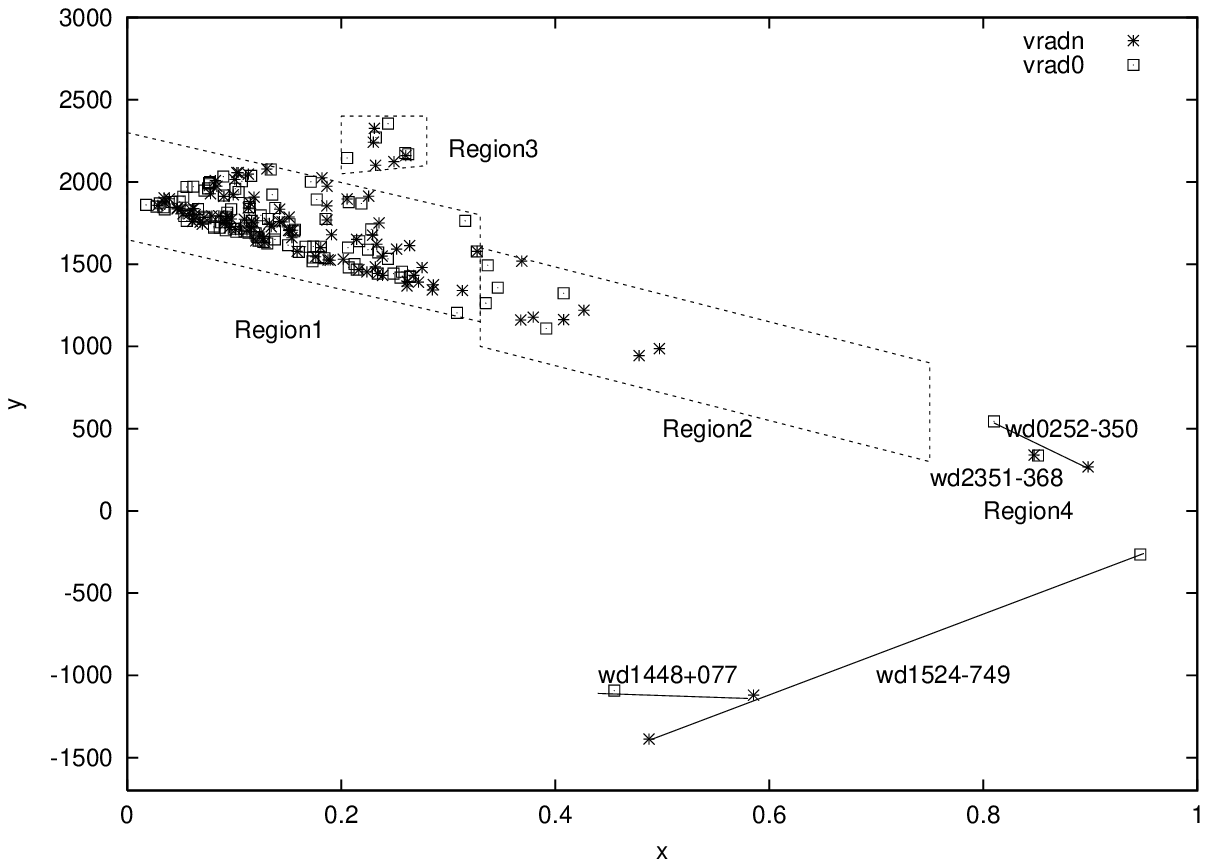}
  \end{psfrags}
  \caption{Effect of setting $v_{\rm rad}=0$ in the $Jz$-$e$-diagram, the change in position of the halo white dwarfs is indicated by lines}
  \label{eccjz0}
\end{figure*}

First we compare the $U$\/-$V$-velocity diagrams. In Fig.~\ref{uvwd0} 
the values for $v_{\rm rad} \ne 0$ are indicated 
by asterisks and 
those for $v_{\rm rad}=0$ by open squares. 
As can be seen setting $v_{\rm rad}=0$ generally 
increases both the $U$ and $V$ components. 
The number of white dwarfs outside the $2\sigma$-limit 
is reduced from $14$ (for $v_{\rm rad } \ne 0$) 
to twelve (for $v_{\rm rad}=0$). 
Four stars among them are far outside the $2\sigma$-limit. 
Those are the halo white dwarfs WD\,0252$-$350, WD\,2351$-$368, WD\,1448+077 
and WD\,1524$-$749. 
The star that is affected most is WD\,1524$-$749. 
Its value of $V$ changes from $-167\,{\rm km\,s^{-1}}$ to 
$-35\,{\rm km\,s^{-1}}$. This is not astonishing, as WD\,1524$-$749 has a high 
radial velocity.   
The white dwarf WD\,0252$-$350 is shifted closer to the $2\sigma$-border 
in $V$-direction. 

Secondly we consider the $J_Z$-$e$-diagram (Fig.~\ref{eccjz0}).
All four halo white dwarfs remain in Region~4. 
But the eccentricity of WD\,0252$-$350 is reduced from $0.9$ to $0.81$ 
so that it gets closer to Region~2. 
WD\,1524$-$749 shows a large shift as well, its 
eccentricity increases from $0.49$ to $0.95$ and its value of $J_Z$ 
increases. This indicates that the orbit is changed from retrograde to  
prograde. 
For the other regions we note that 
the eccentricity decreases for nearly all stars. 
In consequence only five of the seven white dwarfs 
originally in Region~2 stay there whereas two are shifted to Region~1. 
Instead of five only four white dwarfs can now be found in Region~3.  

Adopting the definition of a thick disk candidate 
(situated outside the $2\sigma$-limit in the $U$\/-$V$-diagram 
or in Region~2 or 3 in the $J_Z$-$e$-diagram)  
now twelve white dwarfs qualify 
whereas it have been 13 for for $v_{\rm rad } \ne 0$.  

\begin{figure}
  \begin{psfrags}
    \psfrag{rho/kpc}{$\rho/{\rm kpc}$}
    \psfrag{Z/kpc}{$z/{\rm kpc}$}
    \resizebox{\hsize}{!}{
    \includegraphics{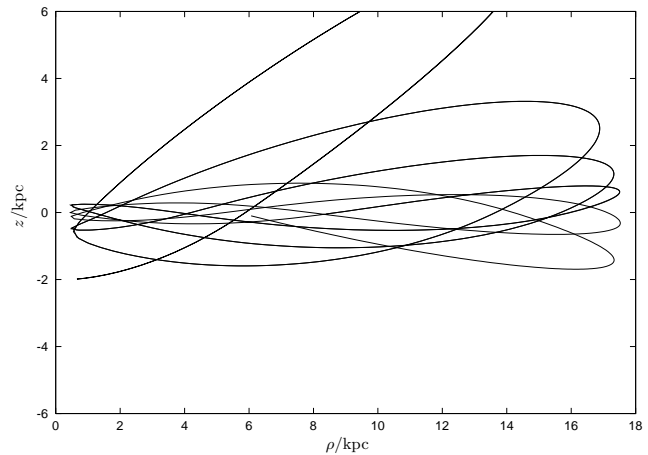}
      }
  \end{psfrags}
  \caption{Orbit of WD\,1524$-$749 for $v_{\rm rad}=0$}
  \label{wd1524_0}
\end{figure}

\begin{figure}
    \begin{psfrags}
      \psfrag{rho/kpc}{$\rho/{\rm kpc}$}
      \psfrag{Z/kpc}{$z/{\rm kpc}$}
  \resizebox{\hsize}{!}{
    \includegraphics{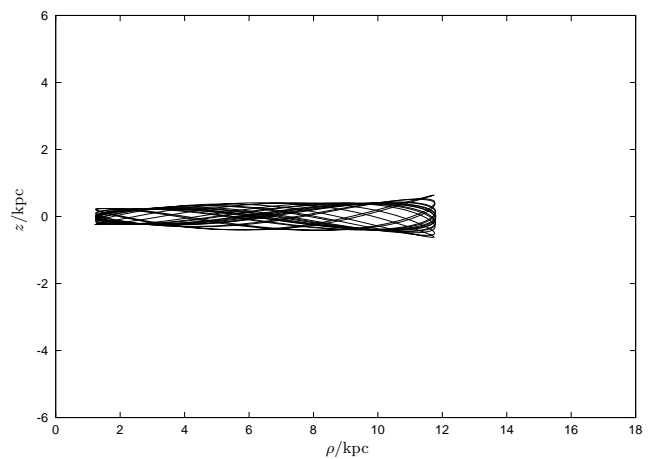}
      }
    \end{psfrags}
  \caption{Orbit of WD\,0252$-$350 for $v_{\rm rad}=0$}
  \label{wd0252_0}
\end{figure}

Last we investigate the effect on the orbits. 
Again the former halo stars are affected strongly. 
The orbit of WD\,1524$-$749 (Fig.~\ref{wd1524_0}) changes from 
a retrograde halo orbit to a typical halo orbit with very high 
extensions in $\rho$ and $Z$. 
The shape of the orbit of WD\,0252$-$350 (Fig.~\ref{wd0252_0}) is 
also altered. 
On the other hand for the thick disk candidates the effects on the orbits 
are small. Only for one white dwarf the orbit changes from thick disk to 
thin disk like.  

Applying our classification scheme in Sect.~\ref{results} 
to the twelve thick disk candidates, ten 
of them pass the criteria for thick disk stars. 
Five with $c \le 2$ are classified as bona fide 
and five with $c=1$ as probable thick disk white dwarfs. 
The fact that we do only loose two of the twelve candidates to the thin disk 
is due to the robustness of the orbits to setting 
$v_{\rm rad}=0$. 
However the number of white dwarfs passing the criteria for 
thick disk candidates is reduced from 13 to ten stars, 
which is a loss of $23\%$.   
Though the eccentricities and the orbits of some halo members 
are altered they are still identified correctly as halo stars. 

\citet{silvestri02} stated that setting $v_{\rm rad}=0$ 
has only very small effects on the kinematics of white dwarfs. 
That may be true compared 
to setting $W=0\,{\rm km\,s^{-1}}$ but we have demonstrated here that the 
effects are not negligible. 
Generally setting $v_{\rm rad}=0$ can lead to an 
underestimate of the fraction of thick disk white dwarfs. 
Misidentifications between thin and thick disk stars can occur 
and halo white dwarfs on retrograde orbits can appear to be on 
prograde ones. 
This should be kept in mind when dealing with samples for which 
radial velocities are not available. 
\begin{table*}
\caption[]
{Mean value and standard deviation of $U$, $V$, 
$W$ for the twelve thick disk 
white dwarfs from our sample,\\$\sigma(U)$, $\sigma(V)$ and $\sigma(W)$
from \citet{chiba00} are shown for comparison\\ \label{uvwtab}}
\begin{tabular}{lcccccc}
\hline 
 & $\left<U\right> $ & $\sigma(U)$ & $\left<V\right>$& $\sigma(V)$ & $\left<W\right>$ & $\sigma(W)$\\
 &${\rm km\,s^{-1}}$ & ${\rm km\,s^{-1}}$ & ${\rm km\,s^{-1}}$ & 
${\rm km\,s^{-1}}$ & ${\rm km\,s^{-1}}$ & ${\rm km\,s^{-1}}$\\
\hline 
Thick disk whited dwarfs from our sample&$-5$ & $69$ & $180$ & $58$ & $-17$ & $43$\\
\hline
Thick disk stars from \citet{chiba00}& &$46$& &$50$& &$35$\\
\hline
\end{tabular}
\end{table*}   
\begin{table*}
\caption[]
{Effective temperatures, surface gravities and masses of the thick disk white dwarf candidates, $M_{\rm W}$, $M_{\rm B}$ denote masses derived from \citet{wood95} and \citet{bloecker97}, respectively\label{thictab}}
\begin{tabular}{lcclcc}
\hline
star & spectra & $T_{\rm eff}$ & ${\rm log}~g$ & $M_{\rm W}$ & $M_{\rm B}$ \\ 
 &  &${\rm K}$ & ${\rm cm \,s^{-2}} $  & \Msolar & \Msolar \\
\hline
HE\,1124+0144 & $ 2$ & $15876$ & $7.685$ & $0.455$ & $0.468$\\
HE\,0409$-$5154 & $ 1$ & $26439$ & $7.750$ & $0.520$ & $0.517$\\
WD\,0013$-$241 & $ 2$ & $18328$ & $7.855$ & $0.545$ & $0.536$\\
WD\,0204$-$233 & $ 1$ & $13176$ & $7.750$ & $0.480$ & $0.483$\\
WD\,0216+143 & $ 1$ & $27132$ & $7.790$ & $0.540$ & $0.530$\\
WD\,0951$-$155 & $ 2$ & $16963$ & $7.765$ & $0.495$ & $0.499$\\
WD\,1334$-$678 & $ 1$ & $ 8958$ & $8.110$ & $0.670$ & $0.649$\\
WD\,1426$-$276 & $ 2$ & $17525$ & $7.665$ & $0.455$ & $0.467$\\
WD\,1515$-$164 & $ 1$ & $13927$ & $7.810$ & $0.510$ & $0.507$\\
WD\,1834$-$781 & $ 2$ & $17564$ & $7.760$ & $0.495$ & $0.498$\\
WD\,1952$-$206 & $ 2$ & $13741$ & $7.780$ & $0.490$ & $0.496$\\
WD\,2322$-$181 & $ 2$ & $21478$ & $7.880$ & $0.565$ & $0.559$\\
\hline
\end{tabular}
\end{table*}
\begin{table*}
\caption[]
{Effective temperatures, surface gravities and masses of the halo white dwarf candidates, $M_{\rm W}$, $M_{\rm B}$ denote masses derived from \citet{wood95} and \citet{bloecker97}, respectively\label{halotab}}
\begin{tabular}{lcclcc}
\hline
star & spectra & $T_{\rm eff}$ & ${\rm log}~g$ & $M_{\rm W}$ & $M_{\rm B}$ \\ 
 &  &${\rm K}$ & ${\rm cm \,s^{-2}} $  & \Msolar & \Msolar \\
\hline
WD\,0252$-$350 & $ 2$ & $17055$ & $7.420$ & $0.350$ & $0.386$\\
WD\,2351$-$368 & $ 2$ & $14567$ & $7.810$ & $0.510$ & $0.509$\\
WD\,1448+077 & $ 1$ & $14459$ & $7.660$ & $0.440$ & $0.455$\\
WD\,1524$-$749 & $ 2$ & $23414$ & $7.605$ & $0.450$ & $0.471$\\
\hline
\end{tabular}
\end{table*} 
\section{Discussion\label{dis}}
We have applied new sophisticated methods to kinematically 
analyse a sample of 107 DA white dwarfs from the SPY project. 
An error propagation method based 
on a Monte-Carlo technique was presented. 
It has been demonstrated that considering only the classical $U$\/-$V$-diagram 
is not sufficient for deciding on population membership. 
In addition it is necessary to take the $J_Z$-$e$-diagram and Galactic orbits
 into account. 
A quantitative method was developed for combining the three 
different classification criteria for thick disk white dwarfs. 
With a consistency check we have demonstrated that our detection 
rate for thick disk stars is very high (93\%) with 
a contamination rate by thin disk stars of only 12\%. 

Four of our white dwarfs prove to be halo members, i.e.\ a fraction 
of 4\%. Two of them are on retrograde orbits. 
This percentage compares well with the results of \citet{sion88}, 
who identified about $5$\% of their sample as halo white dwarfs. 
\citet{liebert89}, on the other hand obtained a percentage of 
14\% halo white dwarfs  
by classifying all stars that exceed a certain value of tangential velocity
as halo members. 
When comparing those samples with ours it has to be kept in mind that 
our selection criteria are finer and permit a separation of thick disk 
and halo. Therefore a part of the white dwarfs classified as halo by 
\citet{sion88} and \citet{liebert89} could belong to the thick disk instead.
Furthermore  both samples suffer from the lack of radial velocities. 

Our sample does not confirm the results of \citet{oppenheimer01} who 
claim that halo white dwarfs constitute an important fraction of the 
dark matter in the Galaxy. 

However, it has to be taken into account that our sample is biased towards 
high temperatures, whereas \citet{oppenheimer01} analyse cool white dwarfs. 
The first set of objects observed by SPY and analysed by \citet{koester01} 
was dominated by relatively hot white dwarfs (mean temperature of the 
\citet{koester01} sample: $21\,700\,{\rm K}$) selected 
from the Hamburg/ESO survey 
\citep{christlieb01} and the white dwarf catalogue of \citet{mccook99}. 

Nine white dwarfs in our sample have been classified as bona-fide thick disk 
and three as probable thick disk. The combined total corresponds to a 
fraction of $11$\% thick disk white dwarfs in our local sample. 

This value is compatible with the one of \citet{silvestri02}. 
Our fraction of $11$\% thick disk white dwarfs, however, is somewhat 
smaller than that of Fuhrmann (2000) (http://www.xray.mpe.mpg.de/fuhrmann/) 
who predicted a fraction of $17$\% thick disk white dwarfs. 
The differences are possibly caused by the bias mentioned above. 
We cannot state an over-representation of white dwarfs compared to 
low mass main-sequence stars which would require a truncated initial 
mass function as suggested by \citet{favata97}.

To discuss the kinematic parameters of our twelve thick disk white 
dwarfs we calculate the mean value and standard 
deviation of the three velocity components.
The results can be found in Table~\ref{uvwtab}. 
Those values are larger than those from \citet{chiba00} who 
analysed the kinematics of 1203 solar-neighbourhood stars with 
metal abundances $[{\rm Fe/H}] \le -0.6$. 
This difference may be due 
to our stringent selection criteria. 
The mean eccentricity has a value 
of $\left<e\right>=0.35$, the mean $z$-component of angular momentum 
is $\left<J_{z}\right>=1\,532\,{\rm kpc\,km\,s^{-1}}$. 

Other interesting issues are the masses and cooling ages of our halo 
and thick disk white dwarfs.  
Masses of the white dwarfs are derived from 
effective temperatures and surface gravities by \citet{koester01} 
using the mass-radius relation from \citet{wood95}'s white dwarf cooling 
tracks. 
In addition we calculated masses with another mass-radius relation 
from \citet{bloecker97}, which is based on full evolutionary 
calculations. 

Results are listed in Table~\ref{thictab} (thick disk white dwarfs) and 
Table~\ref{halotab} (halo white dwarfs) for 
both mass-radius relations. 
$M_W$, $M_B$ denote masses derived from \citet{wood95} and 
\citet{bloecker97}, respectively. 

In the second column the number of spectra available is listed because we
cannot rule out the possibility that stars without a second spectrum are 
spectroscopic binaries. 
All but one star in the halo/thick disk samples have 
$T_{\rm eff}$ above $13\,000\,{\rm K}$ and are thus relatively hot 
and have a low cooling age ($<1\,{\rm Gyr}$).  
Their typical mass is around $0.5$\,\Msolar. 
Only the halo star WD\,0252$-$350 has a rather low mass indicating 
that it does not possess a CO- but a He-core instead.  
Generally our halo and thick disk white dwarfs are hot and have low masses. 
Exception to the rule is the thick disk star, WD\,1334$-$678, which 
is cool ($T_{\rm eff}=8\,958\,{\rm K}$) 
and more massive, $0.61$\,\Msolar. 
The low masses and high temperatures of the majority of the halo and 
thick disk white dwarfs are in line with the assumption that 
these white dwarfs evolved from an old  population of 
long-lived low mass stars. 
Their precursors 
did not yet have time to cool down to low temperatures. 
Since WD\,1334$-$678 is more massive and cooler its origin may be 
different to that of the other thick disk stars. 
We speculate that this star might have been born in a binary system in the 
thin disk and been ejected thereafter from it (run-away star). 
A possible ejection mechanism would be e.g.\ a supernova explosion of  
the primary releasing the secondary at high velocity \citep{davies02}.

The question whether thick disk white dwarfs contribute 
significantly to the total mass of the Galaxy is of great importance 
to clarify the dark matter problem. 
We can estimate this contribution from our results derived above. 
To derive the densities of thin disk and thick disk white dwarfs 
we use the $1/V_{\rm max}$ method \citep{schmidt68}. 
First we calculate the maximum distance $d_{\rm max}$ at which a 
white dwarf would still be included in our sample corresponding 
to $B_{\rm max}=16.5$ for the stars with given $B$-magnitude or 
$V_{\rm max}=16.5$ for the stars with given $V$-magnitude in 
\citet{koester01}. 
The maximum, density-weighted volume $V_{\rm max}$ is 
then computed according to the following equation \citep{green80}: 
\begin{equation}
V_{\rm max}=\int_{0}^{d_{\rm max}}{{\rm e}^{-{l \over l_{0}}}\,{\rm e}^{-{h \over h_{0}}}\,{\rm d}V}   
\label{eq1}
\end{equation}
where $l_{0}$ and $h_{0}$ are the scale lengths and heights for the 
white dwarfs classified as thin and thick disk members, respectively. 
For the thick disk we adopt the values of 
\citet{ojha01}: scale length $l_{\rm 0,thick}=3.7\,{\rm kpc}$, 
scale height $h_{\rm 0,thick}=0.86\,{\rm kpc}$. 
For the thin disk the corresponding scale length is 
$l_{\rm 0,thin}=2.8\,{\rm kpc}$ \citep{ojha01} and the scale height is 
$h_{\rm 0,thin}=0.25\,{\rm kpc}$ \citep{kuijken89,kroupa92,haywood97}. 
As over half of our sample stars possess distances exceeding 
$100\,{\rm pc}$ the assumption that $d$ is small compared to the scale height 
is not valid. Thus $V_{\rm max}$ cannot be computed by the simple 
equation $V_{\rm max}={4\over 3}\,\pi\,d_{\rm max}^3$. 

To check the uniformity of the sample the mean value 
$\left<V \over V_{\rm max} \right>$ is calculated. 
It is found to be $0.4$ which is less than the $0.5$ expected for a 
uniform, spherical distribution. Nevertheless, even a complete sample, 
if selected from a population with a nonuniform space-distribution 
such a disk system, will produce $\left<V \over V_{\rm max} \right><0.5$ 
\citep{green80}.
For a disk-like structure with small scale height
we evaluated $\left<V \over V_{\rm max} \right>=0.4$. 

The white dwarf density of a given population is: 
\begin{equation}
\rho=\sum_{i}{1 \over V_{\rm max,i}}\ .   
\label{eq2}
\end{equation}
For the thick disk density we sum over all 12 thick disk stars, for 
the thin disk density over all 91 thin disk stars. 
To obtain the true density this sum would still have to be multiplied 
by the fraction of the sky $\Omega$ covered by our observations.  
However, it is not necessary to know $\Omega$, 
which is difficult to estimate in our case since 
we are only interested in the fraction of mass contained in 
thick disk white dwarfs versus thin disk white dwarfs and $\Omega$ cancels 
out in this fraction. 

In following we describe how to obtain the mass fraction from 
the densities.
The probability to find a star in a given volume is:
\begin{equation}
p(\vec{r})=C\,{\rm e}^{-{l \over l_{0}}}\,{\rm e}^{-{h \over h_{0}}}\ .   
\label{eq3}
\end{equation} 
We find $C$ by the condition that the probability has to be normalised: 
\begin{equation}
\int_{0}^{\infty}{p(\vec{r})\,{\rm d}V}\stackrel!=1\,\Rightarrow\,
C={{1} \over {\int_{0}^{\infty}{{\rm e}^{-{l \over l_{0}}}\,{\rm e}^{-{h \over h_{0}}}\,{\rm d}V}}}\ .
\label{eq4}
\end{equation} 
The connection between $V_{\rm max }$ and $p(\vec{r})$ is 
obtained by multiplying $V_{\rm max}$ with $C$: 
\begin{equation}
C \cdot V_{\rm max}= \int_{0}^{d_{\rm max}}{p(\vec{r})\,{\rm d}V}\ .
\label{eq6}
\end{equation}
The number density $N(\vec{r})$ is given by $N(\vec{r})=N_{0}\,p(\vec{r})$.

We now consider that we have one star $i$ in the volume $V_{\rm max,i}$. 
\begin{eqnarray}
1 & \stackrel != & \int_{0}^{d_{\rm max}}{N(\vec{r})\,{\rm d}V} \nonumber \\
& = & \int_{0}^{d_{\rm max}}{N_{0,i}\,p(\vec{r})\,{\rm d}V} \nonumber \\ 
& = & \int_{0}^{d_{\rm max}}{N_{0,i}\,C\,{\rm e}^{-{l \over l_{0}}}\,{\rm e}^{-{h \over h_{0}}}\,{\rm d}V}  
\label{eq6}
\end{eqnarray}
\begin{equation}
\Rightarrow N_{0,i}={1 \over {C\,V_{\rm max,i}}}\ .
\label{eq6b}
\end{equation}
By summing over all twelve thick disk white dwarfs the number of thick disk 
stars in the whole Galaxy can be extrapolated: 
\begin{eqnarray}
N_{\rm 0,thick} &  = & \sum_{i=1}^{12}{N_{0,i}} \nonumber \\
&  = & {1\over C_{\rm thick}}\,\sum_{i}\,{1 \over V_{\rm max,i}} \nonumber \\
& = & \int_{0}^{\infty}{{\rm e}^{-{l \over l_{\rm 0,thick}}}\,{\rm e}^{-{h \over h_{\rm 0,thick}}}\,{\rm d}V}\,\sum_{i}{{1 \over V_{\rm max,i}}}  \ . 
\label{eq7}
\end{eqnarray}
The same can be done for the $91$ thin disk white dwarfs. 
Finally we obtain the mass fraction of thick disk versus thin disk 
white dwarfs by  
\begin{equation}
{{M_{\rm thick}\over M_{\rm thin}}}=  
{{N_{\rm 0,thick}\,<M_{\rm thick}>} \over 
{N_{\rm 0,thin}\,<M_{\rm thin}>}} 
\label{eq8}
\end{equation}
with $<M_{\rm thick}>$ and $<M_{\rm thin}>$ being the mean 
mass of the thick and thin disk white dwarfs, respectively. 
Assuming $<M_{\rm thick}>=<M_{\rm thin}>$ yields 
${M_{\rm thick}\over M_{\rm thin}}=0.15 \pm 0.32$. 
The error of $216\%$ was derived using linear error propagation. 
Its large value results from the poor statistics of the 
relatively small thick disk sample. 
Nevertheless we can conclude within the $1\sigma$-level that the total mass of 
thick disk white dwarfs is less than one half 
of the total mass of thin disk white dwarfs. 
Therefore the mass contribution of the thick disk white dwarfs 
cannot be neglected, but it is not sufficient to account for the 
missing dark matter. 
\section{Conclusions\label{con}} 
We have demonstrated how a combination of sophisticated kinematic analysis 
tools can provide a distinction of halo, thick disk and thin disk white 
dwarfs. We have identified a fraction of 4\% halo and 11\% thick disk white 
dwarfs. 
No indications for the existence of a white dwarf dark halo or a 
truncated thick disk IMF function have been found.
Most of our thick disk and halo white dwarfs 
are hot and possess low masses. 
Our results do suggest that the mass present in halo and thick disk 
white dwarfs is not sufficient to explain the missing mass of the Galaxy. 
But to draw definite conclusions more data is needed.  
Our goal is to extend this kinematic analysis to all 
$1\,000$ white dwarfs from the SPY project 
in order to have a large data base to 
decide on population membership of white dwarfs 
and their implications for the mass and evolution of the Galaxy.
\acknowledgements
{
E.-M. P. acknowledges support by the Deutsche Forschungsgemeinschaft 
(grant Na\,365/2-1). E.-M. P. also wishes to express gratitude to the 
Studienstiftung des Deutschen Volkes for a grant. 
Without their travel support it would not have been possible 
to attend the White Dwarf Workshop 2002. 
Thanks go to J. Pauli for interesting and fruitful discussions. 
M. Altmann acknowledges support from the DLR~50~QD~0102. 
We are also grateful for DSS images 
based on photographic data obtained from the UK Schmidt Telescope. 
The UK Schmidt Telescope was operated by the Royal Observatory Edinburgh, 
with funding from the UK Science and Engineering Research Council, until 1988 June, 
and thereafter by the Anglo-Australian Observatory. 
Original plate material is copyright of the Royal Observatory Edinburgh and the Anglo-Australian 
Observatory. 
The plates were processed into the present compressed digital form with their permission. 
The Digitized Sky Survey 
was produced at the Space Telescope Science Institute under US Government grant NAG W-2166.
}
\bibliographystyle{aa}
\bibliography{3220}

\begin{thebibliography}{49}
\expandafter\ifx\csname natexlab\endcsname\relax\def\natexlab#1{#1}\fi

\bibitem[{Alcock {et~al.}(2000)Alcock, Allsmann, Alves, Axelrod, Becker, \&
  Bennett}]{alcock00}
Alcock, C., Allsmann, R., Alves, D.~R., {et~al.} 2000, ApJ, 542, 227

\bibitem[{Allen \& Santillan(1991)}]{allen91}
Allen, C. \& Santillan, A. 1991, Rev. Mex. Astron. Astrofis., 22, 255

\bibitem[{Altmann(2002)}]{altmann02}
Altmann, M. 2002, PhD thesis, Universit\"at Bonn

\bibitem[{Barbier-Brossat {et~al.}(1994)Barbier-Brossat, Petit, \&
  Figon}]{barbier94}
Barbier-Brossat, M., Petit, M., \& Figon, P. 1994, A\&AS, 108, 603

\bibitem[{Bertin \& Arnouts(1996)}]{bertin96}
Bertin, E. \& Arnouts, S. 1996, A\&AS, 117, 393

\bibitem[{Bl\"ocker {et~al.}(1997)Bl\"ocker, Herwig, Driebe, Bramkamp, \&
  Sch\"onberner}]{bloecker97}
Bl\"ocker, T., Herwig, F., Driebe, T., Bramkamp, H., \& Sch\"onberner, D. 1997,
  in Proceedings of the 10th European Workshop on White Dwarfs, ed. J.~Isern,
  M.~Hernanz, \& E.~Garcia-Berro, Astrophysics and Space Science Library, Vol.
  214 (Dordrecht: Kluwer Academic Publishers), p. 57

\bibitem[{Chiba \& Beers(2000)}]{chiba00}
Chiba, M. \& Beers, T.~C. 2000, AJ, 119, 2843

\bibitem[{Christlieb {et~al.}(2001)Christlieb, Wisotzki, Reimers, Homeier,
  Koester, \& Heber}]{christlieb01}
Christlieb, N., Wisotzki, L., Reimers, D., {et~al.} 2001, A\&A, 366, 898

\bibitem[{Davies {et~al.}(2002)Davies, King, \& Ritter}]{davies02}
Davies, M.~B., King, A., \& Ritter, H. 2002, MNRAS, 333, 463

\bibitem[{Dehnen \& Binney(1998)}]{dehnen98}
Dehnen, W. \& Binney, J.~J. 1998, MNRAS, 298, 387

\bibitem[{Edvardsson {et~al.}(1993)Edvardsson, Andersen, Gustafsson, Lambert,
  Nissen, \& Tomkin}]{edvard93}
Edvardsson, B., Andersen, J., Gustafsson, B., {et~al.} 1993, A\&A, 275, 101

\bibitem[{Favata {et~al.}(1997)Favata, Micela, \& Sciortino}]{favata97}
Favata, F., Micela, G., \& Sciortino, S. 1997, A\&A, 323, 809

\bibitem[{{Flynn} {et~al.}(1996){Flynn}, {Sommer-Larsen}, \&
  {Christensen}}]{flynn96}
{Flynn}, C., {Sommer-Larsen}, J., \& {Christensen}, P.~R. 1996, \mnras, 281,
  1027

\bibitem[{Fuhrmann(1998)}]{fuhrmann98}
Fuhrmann, K. 1998, A\&A, 338, 161

\bibitem[{Garcia-Berro {et~al.}(1999)Garcia-Berro, Torres, \& Isern}]{garcia99}
Garcia-Berro, E., Torres, S., \& Isern, J. 1999, in Proceedings of the 11th
  European Workshop on White Dwarfs, ed. J.-E. Solheim \& E.~G. Meistas, ASP
  Conf. Ser. 169, Astronomical Society of the Pacific, p. 69

\bibitem[{Geffert {et~al.}(1997)Geffert, Klemola, Hiesgen, \&
  Schmoll}]{geffert97}
Geffert, M., Klemola, A.~R., Hiesgen, M., \& Schmoll, J. 1997, A\&AS, 124, 357

\bibitem[{Gershenfeld(1999)}]{gershenfeld99}
Gershenfeld, N. 1999, The Nature of Mathematical Modelling (Cambridge:
  Cambridge University Press)

\bibitem[{Green(1980)}]{green80}
Green, R.~F. 1980, ApJ, 238, 685

\bibitem[{Green {et~al.}(1986)Green, Schmidt, \& Liebert}]{green86}
Green, R.~F., Schmidt, M., \& Liebert, J. 1986, ApJS, 61, 305

\bibitem[{{Hagen} {et~al.}(1995){Hagen}, {Groote}, {Engels}, \&
  {Reimers}}]{hagen95}
{Hagen}, H.-J., {Groote}, D., {Engels}, D., \& {Reimers}, D. 1995, \aaps, 111,
  195

\bibitem[{Haywood {et~al.}(1997)Haywood, Robin, \& Creze}]{haywood97}
Haywood, M., Robin, A.~C., \& Creze, M. 1997, A\&A, 320, 440

\bibitem[{{Homeier} {et~al.}(1998){Homeier}, {Koester}, {Hagen}, {Jordan},
  {Heber}, {Engels}, {Reimers}, \& {Dreizler}}]{homeier98}
{Homeier}, D., {Koester}, D., {Hagen}, H.-J., {et~al.} 1998, \aap, 338, 563

\bibitem[{Kerr \& Lynden-Bell(1986)}]{kerr86}
Kerr, F.~J. \& Lynden-Bell, D. 1986, MNRAS, 221, 1023

\bibitem[{{Kilkenny} {et~al.}(1997){Kilkenny}, {O'Donoghue}, {Koen}, {Stobie},
  \& {Chen}}]{kilkenny97}
{Kilkenny}, D., {O'Donoghue}, D., {Koen}, C., {Stobie}, R.~S., \& {Chen}, A.
  1997, \mnras, 287, 867

\bibitem[{Koester {et~al.}(2001)Koester, Napiwotzki, Christlieb, Drechsel,
  Hagen, Heber, \& Homeier}]{koester01}
Koester, D., Napiwotzki, R., Christlieb, N., {et~al.} 2001, A\&A, 378, 556

\bibitem[{Kroupa(1992)}]{kroupa92}
Kroupa, P. 1992, in Complementary approaches to Double and Multiple Star
  Research, ed. H.~A. McAlister \& W.~I. Hartkopf, ASP Conf. Ser. 32,
  Astronomical Society of the Pacific, p. 228

\bibitem[{Kuijken \& Gilmore(1989)}]{kuijken89}
Kuijken, K. \& Gilmore, G. 1989, MNRAS, 239, 571

\bibitem[{{Lamontagne} {et~al.}(2000){Lamontagne}, {Demers}, {Wesemael},
  {Fontaine}, \& {Irwin}}]{lamontagne00}
{Lamontagne}, R., {Demers}, S., {Wesemael}, F., {Fontaine}, G., \& {Irwin},
  M.~J. 2000, \aj, 119, 241

\bibitem[{Liebert {et~al.}(1989)Liebert, Dahn, \& Monet}]{liebert89}
Liebert, J., Dahn, C.~C., \& Monet, D.~G. 1989, in Proceedings of IAU
  Colloquium 114th (Hanover, NH: Springer-Verlag), p. 15

\bibitem[{Luyten(1979)}]{luyten79}
Luyten, W.~J. 1979, LHS catalogue (Minneapolis: University of Minnesota Press)

\bibitem[{McCook \& Sion(1987)}]{mccook87}
McCook, G.~P. \& Sion, E.~M. 1987, ApJS, 65, 603

\bibitem[{McCook \& Sion(1999)}]{mccook99}
---. 1999, ApJS, 121, 1

\bibitem[{Monet {et~al.}(1998)Monet, Bird, Canzian, Dahn, Guetter, \&
  Harris}]{monet98}
Monet, D., Bird, A., Canzian, B., {et~al.} 1998, VizieR On-line Catalog:I/252

\bibitem[{Napiwotzki {et~al.}(2001)Napiwotzki, Christlieb, Drechsel, Hagen,
  Heber, Homeier, \& Karl}]{napiwotzki01}
Napiwotzki, R., Christlieb, N., Drechsel, H., {et~al.} 2001, AN, 121, 503

\bibitem[{Odenkirchen \& Brosche(1992)}]{odenkirchen92}
Odenkirchen, M. \& Brosche, P. 1992, AN, 313, 69

\bibitem[{Ojha(2001)}]{ojha01}
Ojha, D.~K. 2001, MNRAS, 322, 426

\bibitem[{Oppenheimer {et~al.}(2001)Oppenheimer, Hambly, Digby, Hodgkin, \&
  Saumon}]{oppenheimer01}
Oppenheimer, B.~R., Hambly, N.~C., Digby, A.~P., Hodgkin, S.~T., \& Saumon, D.
  2001, Sci, 292, 698

\bibitem[{Ostriker \& Peebles(1973)}]{ostriker73}
Ostriker, J.~P. \& Peebles, P. J.~E. 1973, ApJ, 186, 467

\bibitem[{Reid {et~al.}(2001)Reid, Kailash, \& Hawley}]{reid01}
Reid, I.~N., Kailash, K.~C., \& Hawley, S.~L. 2001, ApJ, 559, 942

\bibitem[{Rubin {et~al.}(1978)Rubin, Ford, \& Thonnard}]{rubin78}
Rubin, V., Ford, W. K.~J., \& Thonnard, N. 1978, ApJ, 225, 107

\bibitem[{{Schmidt}(1968)}]{schmidt68}
{Schmidt}, M. 1968, \apj, 151, 393

\bibitem[{Silvestri {et~al.}(2002)Silvestri, Oswalt, \& Hawley}]{silvestri02}
Silvestri, N.~M., Oswalt, T.~D., \& Hawley, S.~L. 2002, AJ, 124, 1118

\bibitem[{Silvestri {et~al.}(2001)Silvestri, Oswalt, Wood, Smith, Reid, \&
  Sion}]{silvestri01}
Silvestri, N.~M., Oswalt, T.~D., Wood, M.~A., {et~al.} 2001, AJ, 121, 503

\bibitem[{Sion {et~al.}(1988)Sion, Fritz, McMullin, \& Lallo}]{sion88}
Sion, E.~M., Fritz, M.~L., McMullin, J.~P., \& Lallo, M.~D. 1988, AJ, 96, 251

\bibitem[{{Stetson}(1992)}]{stetson92}
{Stetson}, P.~B. 1992, in ASP Conf. Ser. 25: Astronomical Data Analysis
  Software and Systems I, Vol.~1, 297

\bibitem[{{Wisotzki} {et~al.}(2000){Wisotzki}, {Christlieb}, {Bade},
  {Beckmann}, {K{\" o}hler}, {Vanelle}, \& {Reimers}}]{wisotzki00}
{Wisotzki}, L., {Christlieb}, N., {Bade}, N., {et~al.} 2000, \aap, 358, 77

\bibitem[{{Wisotzki} {et~al.}(1996){Wisotzki}, {Koehler}, {Groote}, \&
  {Reimers}}]{wisotzki96}
{Wisotzki}, L., {Koehler}, T., {Groote}, D., \& {Reimers}, D. 1996, \aaps, 115,
  227

\bibitem[{Wood(1995)}]{wood95}
Wood, M.~A. 1995, in White Dwarfs, Proceedings of the 11th European Workshop on
  White Dwarfs Held at Kiel, Germany, 29 August-1 September 1994, ed.
  D.~Koester \& K.~Werner, ASP Conf. Ser. 169, Astronomical Society of the
  Pacific, p. 348

\bibitem[{Zacharias {et~al.}(2000)Zacharias, Urban, Zacharias, Hall, Wycoff, \&
  Rafferty}]{zacharias00}
Zacharias, N., Urban, S.~E., Zacharias, M.~I., {et~al.} 2000, AJ, 120, 2131

\end{thebibliography}
\end{document}